\documentclass[10pt,journal,compsoc]{IEEEtran}
\usepackage{enumitem}
\usepackage[nocompress]{cite}
\usepackage[utf8]{inputenc}
\usepackage[dvipsnames]{xcolor}
\usepackage{graphicx}
\graphicspath{{img/}}
\usepackage{float,wrapfig}
\usepackage[breaklinks=true]{hyperref}
\usepackage{url}
\usepackage{booktabs,tabularx,multirow}
\usepackage{caption, subcaption}
\usepackage[htt]{hyphenat}
\usepackage{amsmath}
\usepackage{amsfonts}
\usepackage{rotating}

\title{An architectural technical debt index based on machine learning and architectural smells}
\author{
    Darius Sas, Paris Avgeriou \IEEEcompsocitemizethanks{
        \IEEEcompsocthanksitem
        Darius Sas and Paris Avgeriou are with the Bernoulli Institute for Mathematics, Computer Science, and Artificial Intelligence, University of Groningen, Groningen, Netherlands 9714GV.
        E-mail: \{d.d.sas,p.avgeriou\}@rug.nl, darius.sas@outlook.com.}}
\date{\today}

\begin{document}
\IEEEtitleabstractindextext{
\begin{abstract}
    A key aspect of technical debt (TD) management is the ability to measure the amount of principal accumulated in a system.
    The current literature contains an array of approaches to estimate TD principal, however, only a few of them focus specifically on architectural TD, but none of them satisfies all three of the following criteria: being fully automated, freely available, and thoroughly validated.
    Moreover, a recent study has shown that many of the current approaches suffer from certain shortcomings, such as relying on hand-picked thresholds.
    
    In this paper, we propose a novel approach to estimate architectural technical debt principal based on machine learning and architectural smells to address such shortcomings.
    Our approach can estimate the amount of technical debt principal generated by a single architectural smell instance.
    To do so, we adopt novel techniques from Information Retrieval to train a learning-to-rank machine learning model (more specifically, a gradient boosting machine) that estimates the severity of an architectural smell and ensure the transparency of the predictions.
    Then, for each instance, we statically analyse the source code to calculate the exact number of lines of code creating the smell.
    Finally, we combine these two values to calculate the technical debt principal.
    
    To validate the approach, we conducted a case study and interviewed 16 practitioners, from both open source and industry, and asked them about their opinions on the TD principal estimations for several smells detected in their projects.
    The results show that for 71\% of instances, practitioners agreed that the estimations provided were \emph{representative} of the effort necessary to refactor the smell.
\end{abstract}
}
\maketitle
\IEEEdisplaynontitleabstractindextext

\section{Introduction}\label{sec:Intro}
The technical debt (TD) metaphor borrows the concepts of \emph{principal} and \emph{interest} from the financial domain and uses them to convey key software maintenance concepts.
In particular, debt principal indicates the effort required to \emph{fix} a current, non-optimal solution, whereas debt interest indicates the recurrent effort necessary to \emph{keep maintaining} it \cite{Avgeriou2016}.
As an example, consider a portfolio management system that requires massive revisions in order to accommodate for the changes required by the customer\cite{Cunningham1992}. The interest represents the recurrent costs of making the revisions, whereas the principal is the cost of completely replacing the solution with a new one that would allow these changes to be seamless.

The importance of managing TD is ever increasing, especially for architectural TD (ATD), as architectural decisions were found to be the greatest source of TD faced by practitioners \cite{Ernst2015}.
A key part of managing TD is to be able to \emph{measure} the amount of TD principal incurred by an application, but this has not yet been effectively addressed in the state of the art.
Theoretically, the problem of measuring the TD principal requires defining a function that transforms maintenance-related data points (metrics, smells, violations of rules or principles, etc.) into a single number representing the overall effort required to fix them.
Over the past years, several studies proposed approaches to estimate the amount of debt principal accrued by an application \cite{Khomyakov2020,Avgeriou2021}, both at the architectural level and at code or design levels; however, most of these studies relied on techniques that have known shortcomings and resulted in estimation functions that were not \emph{thoroughly} validated \cite{Khomyakov2020}.
A common weakness shared by many of these studies is the use of hand-picked thresholds or relying on benchmarks that include arbitrary systems (of arbitrary size, domain, etc.) to determine these thresholds (see Section \ref{sec:related-work} for more details).
Moreover, while some of these approaches are fully automated, they are no freely-available implementations that can be used by others to replicate the results obtained.

In this paper, we propose a novel approach to estimate ATD principal, called \textbf{\emph{ATDI}} (Architectural Technical Debt Index), by adopting machine learning (ML) to overcome the aforementioned shortcomings of existing approaches.
The main advantage of using ML over thresholds or benchmarks is that ML does not require picking these manually; the model will automatically deduce these from the data.

In our approach, we use architectural smells (AS) as the main proxy for measuring ATD.
AS represent decisions that violate design principles and result in undesired dependencies, overblown size, and excessive coupling \cite{Lippert2006,Garcia2009} among the classes and packages of a system.
The main advantage of using AS as a proxy for ATD is that we can estimate the amount of principal each AS contributes to the system while also being easy to detect automatically with high precision (see Section \ref*{sec:AS}). 
This provides a benefit over simply using metrics as proxies for ATD because using AS is: a) \textbf{actionable} as practitioners can make prioritisation decisions using AS; and b) \textbf{targeted} as practitioners know exactly what the problem is, where it is, and how it should be addressed.
The main disadvantage of using AS is that they are but a part of all the possible forms that ATD can assume, therefore it is not guaranteed that all of the ATD principal is represented.
Nevertheless, AS are the most common form of ATD studied in the literature \cite{Verdecchia2018}, they have been recognised as particularly problematic in industry \cite{Fontana2020,Sas2021b}, and  they have been used by other approaches in the literature as proxies for estimating the whole ATD principal \cite{Xiao2016,Roveda2018}. 
Further details on the choice of AS as a threat to validity are discussed in Section \ref{sec:threats-to-validity}.

Our approach, uses ML to calculate the severity of each AS instance (i.e. how harmful it is to Maintainability/Evolvability) and then \emph{combines} it with precise static analysis of the source code to determine the lines of code responsible for the smell (which we use to gauge the size of the smell within the system); this combination is used to calculate the ATD principal as an \emph{index}.
To train the machine learning model (a gradient boosting machine, to be precise), we create a data set using techniques from Information Retrieval to compare AS and rank them by their severity.
The results of the training show that the ML model can successfully rank AS by their severity with a \textbf{high degree of accuracy} by achieving a .97 of $NDCG$ (Normalized Discounted Cumulative Gain \cite{Jarvelin2002}), the de facto standard metric used to evaluate the type of ML model we used in our approach (see Section \ref{sec:perf-metric} for details).
Moreover, to ensure the predicted severity is justified (e.g. not biased by a variable irrelevant to a specific smell) and \textbf{transparent}, we employ a state-of-the-art technique, called SHAP \cite{Strumbelj2014}, to visually analyse a small sample of predictions.
The results show how exactly the considered variables contribute to the predictions.

After ensuring the ML model is predicting severity correctly, we \textbf{validate} the output of the whole approach. 
We inspect whether the estimations provided by our approach are actually relevant to developers by checking whether they are \emph{representative} of the repayment effort perceived and \emph{meaningful} with respect to each other (e.g. this smell requires twice the effort to refactor than this other smell, and has a double ATDI value).
To this end, we interview \textbf{16 practitioners} from both the open source and industrial world.
Each interviewee is shown a number of AS instances in their own systems, as well as the respective ATD principal estimation provided by our approach.
In 71\% of the cases, interviewees totally agree with the estimations provided by our approach and deem them representative of the effort necessary to repay the debt.

This paper's structure is as follows:
Section \ref{sec:AS} summarises the theory of architectural smells and the tool used to detect them; Section \ref{sec:approach} introduces the approach we developed to estimate ATD principal as an index; Section \ref{sec:study-design} elaborates on the case study design, including the data collection and analysis methodologies; Section \ref{sec:descriptive-statistics} presents some descriptive statistics about ATDI; Sections \ref{sec:rq1-results} and \ref{sec:rq2-results}  present the results of the two research questions; Section \ref{sec:discussion} discusses possible implications of the results for researchers and practitioners; Section \ref{sec:threats-to-validity} describes the threats to the validity of this study and how they were mitigated; Section \ref{sec:related-work} lists the related work and compares it with the results obtained by this study; and finally, Section \ref{sec:conclusion-fw} concludes the paper and lists possible future work opportunities.

\section{Architectural smells}\label{sec:AS}
The AS considered in this study are the following 4 types: Cyclic Dependency (CD), Unstable Dependency (UD), Hublike Dependency (HD), and God Component (GC).
The first sub-section provides a brief definition for each type with few details on how they are detected by the used tool; for further details, we refer the reader to Arcelli et al. \cite{Fontana2016}. 
The second and third sub-sections provide respectively a brief description of the tool used to detect architectural smells, called \textsc{Arcan}, and the smell characteristics used to calculate ATDI. 

\subsection{Definition and implications}
Lippert and Roock \cite{Lippert2006} define architectural smells as violations of recognised design principles (such as the ones defined by Martin \cite{Martin2018}) that result in undesired dependencies, overblown size, and excessive coupling \cite{Garcia2009}.
Architectural smells are an indication that something may be problematic, but they do not necessarily imply so.

This definition of architectural smells may sound very similar to the definition of \emph{code smells} provided by Kent Beck\footnote{Read \url{https://wiki.c2.com/?CodeSmell} for more info.}.
However, there is a clear distinction between the two: Architectural smells involve multiple classes, packages, architectural layers, or even sub-systems \cite{Lippert2006}, whereas code smells (CS) arise at line of code, method, or class level \cite{Fowler2002}.
This means that architectural smells, contrary to code smells, require \emph{large refactorings} in order to be removed from the system \cite{Lippert2006}.

It is important to mention that previous work provides empirical evidence that the AS  considered in this study and the most well-known CS are \emph{independent} entities and that there is no correlation between the presence of AS and CS \cite{Fontana2019}.

Both AS and CS manifest themselves in different forms that are commonly referred to as different \emph{types}.
Some examples of CS types are \emph{Duplicated Code}, \emph{Long Method}, and \emph{Large Class} \cite{Fowler2002}.

In this study, we chose to focus on four types of AS: Cyclic Dependency (CD), Hub-Like Dependency (HL), Unstable Dependency (UD), and God Component (GC) \cite{Fontana2016,Lippert2006,Sas2019}. We opted to study these AS because they are some of the most prominent architecture smells, and there already exists tools that support their automatic detection \cite{Fontana2016, Fontana2017}.

\paragraph{Unstable dependency (UD)}\label{sec:arch-smells-ud}
This smell represents a package that depends upon a significant number of packages that are less stable than itself.
The stability of a package is measured using Martin's instability metric \cite{Martin2018}, which measures the degree to which a package is susceptible to change because of its dependencies.
The tool \textsc{Arcan} uses a 30\% threshold \cite{Fontana2017} on the number of packages that are less stable to detect this smell.

The main problem caused by UD is that the probability to change the main package grows higher as the number of unstable packages it depends upon grows accordingly. This increases the likelihood that the packages that depend upon it change as well when it is changed (ripple effect), thus inflating future maintenance efforts.

\paragraph{Hublike dependency (HL)}\label{sec:arch-smells-hl}
This smell represents an artefact (called hub) where the number of ingoing and outgoing dependencies is higher than the median in the system \cite{Fontana2016}. A hublike dependency can be detected both at the package and at the class level.

The implications of this smell for development activities are once again concerning the probability of change and the ease of maintenance.
Making a change to any of the artefacts that the hub depends upon may be very hard \cite{Martin2018} because many other artefacts may indirectly depend on them even though there is only one artefact \emph{directly} depending on them (the hub).
Additionally, the hub is also overloaded with responsibility and has a high coupling.
This structure is thus not desirable, as it increases the potential effort necessary to make changes to all of the elements involved in the smell.

\paragraph{Cyclic dependency (CD)}\label{sec:arch-smells-cd}
This smell represents a dependency cycle among a number of artefacts; there are several software design principles that suggest avoiding creating such cycles \cite{Lippert2006,Parnas1979,Stevens1974,Martin2018}.

Cycles affect mostly complexity, but their presence also has an impact on compiling (causing the recompilation of big parts of the system), testing (forcing to execute unrelated parts of the system, increasing testing complexity), or deploying (forcing developers to re-deploy unchanged components) \cite{Lippert2006}.

\paragraph{God component (GC)}\label{sec:arch-smells-gc}
This smell represents a component (or package, in Java) that is considerably larger in size (i.e. lines of code) than other components in the system \cite{Lippert2006}.
Originally, GC was defined using a fixed threshold on the lines of code \cite{Lippert2006}, \textsc{Arcan} however uses a variable benchmark based on the frequencies of the number of lines of code of the other packages in the system \cite{Fontana2015}.
Adopting a benchmark to derive the detection threshold fits particularly well in this case because what is considered a ``large component'' depends on the size of other components in the system under analysis and in many other systems.
A benchmark allows to make precisely this kind of comparisons.

God components aggregate too many concerns together in a single package and they are generally a sign that there is a missing opportunity for splitting up the package into multiple sub-packages.
God components are often the result of several small incremental changes over a long period of time, sometimes effectively implementing a lot of the overall functionality of the system.
Over time, the understandability of the component deteriorates along with the reusability of the individual parts of the component, as developers are not keen to use a piece of software that is difficult to understand \cite{Lippert2006}.

\subsection{Arcan}
To detect AS, we used a tool called \textsc{Arcan}.
\textsc{Arcan}'s detection capabilities were validated by previous studies and obtained a precision ranging from 70\% to 100\% \cite{Fontana2020, Fontana2017}, depending on the project and type of smell considered.

\textsc{Arcan} parses Java (by relying on Spoon \cite{Pawlak2015}), C, and C++ source code files to create a dependency graph where files, components, classes and packages are all represented using different nodes with different labels. Dependencies, and other relationships between nodes, are represented using edges that connect the dependant to its dependencies with an outgoing, labelled edge (e.g. if artefact \texttt{A} depends on artefact \texttt{B}, then the dependency graph contains a directed edge connecting \texttt{A} to \texttt{B}.). 
The project's structural information contained in the dependency graph is then used to calculate several software metrics (e.g. fan-in, fan-out, instability \cite{Martin2018}, etc.) and then detect architectural smells by recognising their structure in the dependency graph.

Compared to other tools, \textsc{Arcan} uses only software metrics and structural dependencies in order to detect architectural smells. 
This makes \textsc{Arcan} different from tools such as DV8 \cite{Xiao2016} (a tool used by related work) which also requires the use of change metrics.
The command line version of \textsc{Arcan} used for this study is available in the replication package \cite{ReplicationPackage}.

\subsection{Smell characteristics}
An architectural smell \emph{characteristic} is a property or attribute of an architectural smell instance \cite{Sas2019}. 
An architectural smell \emph{instance} is a concrete occurrence of a type of architectural smell.
For each architectural smell type, one can measure different characteristics.
In this work, we are going to use architectural smell characteristics as features (i.e. inputs) for a machine learning model (more details in Section \ref{sec:calculating-severity}).
The characteristics considered in this work are described in Table \ref{tab:characteristics}.

\begin{table}[tbp]
    \footnotesize
    \centering
    \caption{Architectural smell characteristics relevant in this study. \textit{PCT: Package Containment Tree}}\label{tab:characteristics}
    \begin{tabular}{p{0.25\linewidth}|p{0.69\linewidth}}\toprule
        \textbf{Name} & \textbf{Description} \\ \midrule
        Size & The number of artefacts affected by the smell. \\
        Number of edges & The number of dependency edges among the affected artefacts. \\ 
        PageRank & The importance of the artefacts affected by the smell within the dependency network of the system \cite{Roveda2018}. \\
        Affected Type & The type of the affected artefact (i.e. either class or package) \\
        PCT Depth* & Depth refers to the number of packages that are an ancestor of the affected element in the system's package hierarchy (i.e. the PCT) \cite{Laval2012}. \\
        PCT Distance* & The number of packages that need to be traversed in the PCT to reach an affected element of the smell starting from another affected element \cite{Laval2012,Al-Mutawa2014}.\\
        Shape & (for CD only) The shape of a cycle: tiny, circle, chain, star, clique (from \cite{Al-Mutawa2014}). \\
        Instability gap & (for UD only) Is the difference between the instability of the main component and the average instability of the dependencies less stable than the component itself \cite{Fontana2016}. \\\midrule
    \end{tabular}
    \scriptsize{*Since every smell affects multiple elements, and PCT metrics are calculated individually on the classes and packages affected by the smell, we aggregate them as a mean and standard deviation.}
\end{table}

\section{The approach}\label{sec:approach}
This section describes the approach we designed to calculate an architectural technical debt index, or ATDI.
As discussed in Section \ref{sec:Intro}, our approach is based exclusively on architectural smells (AS) and does not consider other types or forms of technical debt. 

\subsection{Indexes and cost estimates}
Theoretically, technical debt (TD) principal is defined as the \emph{cost} necessary to develop a better solution than the currently implemented one \cite{Avgeriou2016}, easing future maintenance and evolution efforts.
Similarly, architectural technical debt (ATD) principal refers to the same concept, but focuses on architectural solutions only.
Several tools, both commercial and open source \cite{Avgeriou2021,Khomyakov2020}, claim to estimate the cost to repay the TD principal of a software system using just source code artefacts.
In practice, however, calculating the exact cost of remediation is a rather ambitious task, as several factors -- both internal and external to the codebase and the company -- may influence it and vary depending on context, organisation and country \cite{Murillo2021,Rios2020,Rios2018}.
If these are not taken into account, the estimate could be imprecise and not reflect the actual cost. 
An index, on the other hand, is not associated with an exact cost, but rather it correlates with the \textbf{\emph{effort}} necessary to remediate the technical debt incurred by the current solution.
It also does not make any assumptions regarding the cost of development, thus avoiding misleading engineers and misrepresenting the actual costs.
Therefore, we opted to treat the ATD principal calculated through our approach as an index, rather than as an estimation of the cost.

To link our approach to the effort needed to refactor, we use static analysis to extrapolate the exact lines of code that create the AS as well as the metrics listed in Table \ref{tab:characteristics}.

The importance of choosing an index over a cost estimate emerged during the design of our approach when we received feedback on the matter from two industrial experts. 
Both experts suggested to avoid a cost estimation as this would spark unnecessary discussion and create controversy and confusion among the developers, architects, and managers who would have different opinions, ultimately leading to distrust against the provided values.
Note that this anecdotal evidence is put to the test by the validation process described in the study design section (Section \ref{sec:study-design}).

To sum up, we do not aim at estimating the \emph{cost impact} of technical debt \cite{Avgeriou2016}, but only the \emph{effort required} to fix the current solution \cite{Avgeriou2016} expressed as an index. 
Using an index over a monetary estimation allows for a more concise and unbiased representation of the effort necessary to remediate the incurred TD. 
Related work from Section \ref{sec:related-work-atd} and Table \ref{tab:rw-comparison} show that this is also a common choice in the literature when estimating ATD principal.

\subsection{Definition}\label{sec:approach-definition}
The simplest and most intuitive way of estimating the ATD index based on AS is by summing up the individual indexes of each smell\cite{Ampatzoglou2018}.
This is the solution adopted by previous studies as well \cite{Letouzey2010,Curtis2012,Marinescu2012,Roveda2018} and (1) allows users to quickly understand the impact of one instance on the overall value of the index, and (2) it resonates with the financial metaphor, where the total amount of debt is the sum of all the debts.
Also note that AS are used as an indicator of the presence of debt, thus acting as a proxy for its estimation.

Formally, we define the ATD principal index as
\begin{equation}\label{eq:atdi}
    ATDI(P) =  \sum_i^{S_P} ATDI(x_i)
\end{equation}
where $x_i$ are the architectural smells $S_P$ detected in the project $P$.
This value can be normalised by the size of the project $P$ in lines of code (LOC) to obtain the density of ATDI per 1000 lines of code, to allow us to compare values obtained from different projects
\begin{equation}\label{eq:atdi-normalised}
    ATDI_{density}(P) = \frac{ATDI(P)}{LOC(P)} \cdot 1000
\end{equation}
The density is expressed for every 1000 LOC in order to reduce the number of decimals in the case of very high values of $LOC(P)$.

The index of a single smell is calculated as the product of: 
\begin{equation}\label{eq:atdi-smell}
    ATDI(x_i) = s(x_i) \cdot m(x_i)
\end{equation}
\begin{enumerate}[label=\alph*)]
    \item \textbf{\emph{Severity}}, calculated by the function $s : S_P \rightarrow  [1, 10]$. \emph{Severity} was used consistently in previous studies to estimate TD principal \cite{Roveda2018,Marinescu2012,Curtis2012}. In our case, we adopted Marinescu's \cite{Marinescu2012} approach to define severity in the range $[1, 10]$ with higher values representing more severe smells\footnote{There is also a more practical reason described in Section \ref{sec:calculating-severity}.};
    \item \textbf{\emph{Extent}}, calculated by the function $m : S_P \rightarrow  \mathbb{N}_{\ge 1}$ and defined as the number of the lines of code that contribute to the creation of the smell, giving an estimation of its size within the system.
    The \emph{extent}, or number of lines of code, was used for estimating the amount of effort in previous studies on technical debt \cite{Chatzigeorgiou2015,Kamei2016,Nugroho2011} and non-technical debt related work as well as a proxy of complexity \cite{Morasca2001, Kitchenham2004}. 
\end{enumerate}

The definition from Equation \ref{eq:atdi-smell} allows us to model the intuition that \emph{more severe smells are more detrimental to maintainability (as well as evolvability)} \cite{Roveda2018} and \emph{more extended smells require more effort to be removed} \cite{Nugroho2011}.
Therefore, ATDI is a proxy for the effort necessary to repay the ATD present in the system affecting both Maintainability and Evolvability.

More details on the information used by previous approaches and existing tools to calculate their indexes are summarised by Avgeriou et al. \cite{Avgeriou2021} and Khomyakov et al. \cite{Khomyakov2020}.
In the following two sub-sections we elaborate on the definition of the two concepts, i.e. severity and extent.

\subsubsection{Defining severity}\label{sec:approach-definition-severity}
In software engineering, the term \emph{severity} is commonly used to describe how harmful a certain type of issue (e.g. architectural smells) is with respect to (w.r.t) a certain quality attribute (e.g. Maintainability or Evolvability).
Severity is used to gauge the impact of different instances of the same type of smell and decide which one is more harmful to the system \cite{Marinescu2012}.

Similarly to the case of code smells \cite{ArcelliFontana2017, ArcelliFontana2015} and design flaws \cite{Marinescu2012}, the severity of an architectural smell is determined by the properties of the structure of the smell instance itself, measured by smell characteristics \cite{Sas2019} (Table \ref*{tab:characteristics}).
For example, assuming all the other characteristics are equal, a cycle with 5 nodes and 5 edges is much easier to refactor than a smell with 5 nodes and 20 edges.

Marinescu \cite{Marinescu2012}, Vidal et al. \cite{Vidal2016}, and Tsantalis et al. \cite{Tsantalis2011} proposed approaches to calculate severity based on a number of metrics related to the flaw in question, including cohesion, coupling, past changes, and complexity.
Our approach is similar as we calculate severity by using architectural smell characteristics \cite{Sas2019} to measure certain properties of a smell instance (and therefore, indirectly, of the artefacts affected).
Smell characteristics were used in previous studies on architectural smells for calculating the ATD index  \cite{Roveda2018}, and to manually determine the severity label for a code smell to be used by machine learning models as well \cite{ArcelliFontana2017}.

\subsubsection{Defining extent}\label{sec:approach-extent}
We define as the \emph{extent} of an an architectural smell the number of lines of code in a source code artefact that break the rules used to detect an architectural smell.
For example, in the case of a cyclic dependency between two files, the lines of code in those two files that are responsible for the dependencies creating the cycle among them.
In general, the purpose is to calculate how extended a smell is within the system in order to gauge the amount of complexity that a developer needs to understand and tackle while applying meaningful changes to the codebase in order to remove the smell.
The LOC metric was consistently used by previous studies as a proxy of complexity \cite{Morasca2001, Kitchenham2004, Morozoff2010}.
The idea behind selecting the extent of a smell as the proxy for the effort necessary to remove the smell is that the more lines of code the smell is made of, the more coupled it is to that specific source code artefact (class, package, etc.).

Practically speaking, we take into account how many lines of code of the system must be changed, or must be taken into consideration (understood), in order to eliminate the smell from the code base.
This approach resembles previous research on the topic \cite{Nugroho2011}, where lines of code where used as a starting point for the estimation of the effort.

\subsection{Calculation of the index}\label{label:index-calculation}
This section details the steps necessary to calculate ATDI (Equation \ref{eq:atdi-smell}) for the architectural smell types we take into consideration in this study.

\subsubsection{Calculating severity}\label{sec:calculating-severity}
The \emph{severity} of an architectural smell depends on several factors, making it hard to derive rules of thumb for determining when a smell is more severe than another.
For example, a Cyclic dependency $A$ between 10 classes may be less severe than a cycle $B$ between 5 packages, despite affecting more elements (i.e. larger size). Yet, another cyclic dependency $C$ affecting 10 classes can be more severe than $B$ if the elements belonging to $C$ cross package boundaries \cite{Laval2012}.
Therefore, just relying on one or more smell characteristics (e.g. size and affected type) is not very helpful.

To calculate severity, we will instead use a specific class of machine learning (ML) models that are able to rank different smell instances in order of their severity.
This class of ML models is typically referred to as learning to rank (LTR) models \cite{TieYan2009}.
A LTR model is trained using a list of documents (e.g. web pages) that have some partial order defined among them (e.g. relevance to a certain query).
The order is typically induced by giving a numerical or ordinal score to each item in the list.
The goal of the LTR model is to produce a permutation of items (i.e. rank them) in new lists (i.e. not part of the training set).
Formally, given a list of architectural smells $X = x_1,x_2,...,x_n$, LTR models try to learn a function $f(X)$ that predicts the relevance (i.e. the \emph{severity} in our case) of any given smell $x_i$.
The relevance is usually a numerical or ordinal score: the higher the value, the more relevant (severe) the smell is.

The main difference of LTR models from traditional classification or regression models, is the training process. An LTR model tries to optimise for the \emph{ranking} of the whole training set, whereas classification and regression models try to minimise the error of the predicted and actual label/value of each entry in the training set. 
Using the terminology from our domain, LTR models try to \textit{\textbf{minimise the number of times a more severe smell is ranked below a less severe smell}}.

LTR models are trained using labels that determine relevance, where higher values imply higher relevance\footnote{See \url{https://lightgbm.readthedocs.io/en/latest/Parameters.html}.}. 
Therefore, our data set will have pairs such as $\langle x_i, s \rangle$, where $x_i$ is the smell and $s\in [1, 10] \subset \mathbb{N}$ is the severity label that our algorithm is trying to learn.
A smell $x_i$ is represented using its characteristics as features, such as the number of elements affected, the number of edges, the page rank in the dependency network of the system, and several others. More details on the training process and creation of the data set are provided in Section \ref{sec:rq1-results}.

\subsubsection{Calculating extent}\label{sec:calculating-extent}
The calculation of the \emph{extent} of an architectural smell depends on the rules used to detect the architectural smell.
For our approach, we are focusing on four types of architectural smells: Cyclic Dependency (CD), Hublike Dependency (HL), Unstable Dependency (UD) and God Component (GC).
The calculation of $m$ from Equation \ref{eq:atdi-smell} for these four smell types translates into two different cases:
\begin{itemize}
    \item for dependency-based smells (CD, HL, UD), we have the number of lines of code generating and using the dependencies between the artefacts taking part in the same smell;
    \item for size-based smells (GC), we have the number of lines of code exceeding the median lines of code of packages/components in the system.
\end{itemize}
The upcoming paragraphs will cover in detail the reasoning behind these choices.

\paragraph{Dependency-based smells}
For CD, HL and UD, the number of lines of code using and generating the dependencies creating the smell has been selected as a proxy to calculate $m$.
Effectively, these are the lines of code contributing to the smell creation (i.e. the dependencies), and therefore they must be taken into consideration during refactoring, thus it can be a indicator for the refactoring effort.
However, this does \emph{not imply} that all dependencies are going to be removed, but the complexity of removing the smell is a \emph{function of the number of lines of code creating and using those dependencies}.

\begin{figure}[]
    \centering
    \includegraphics[width=1\linewidth]{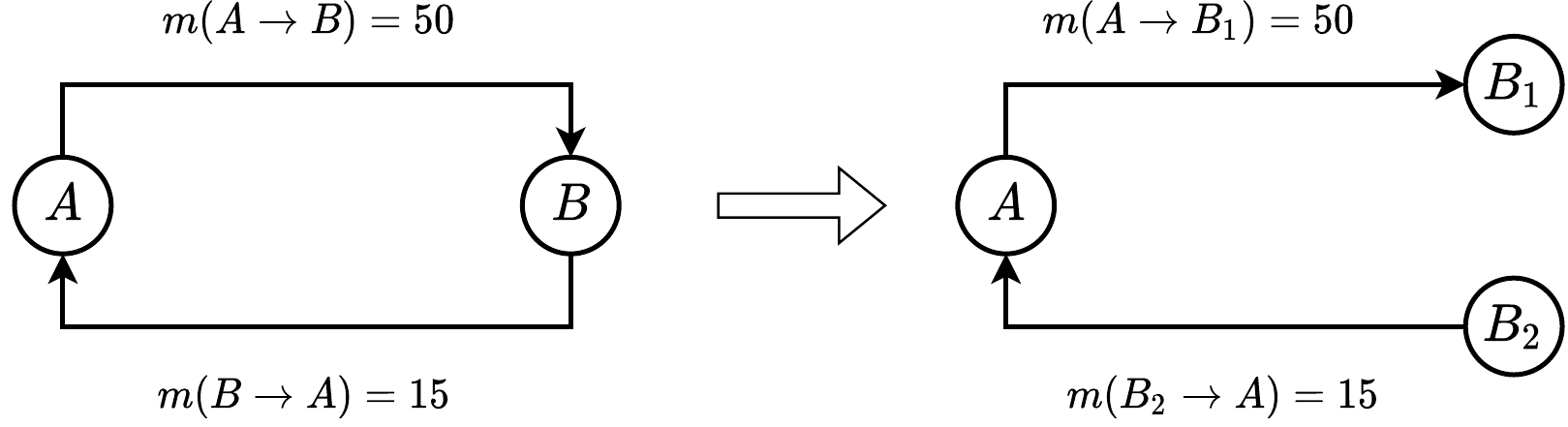}
    \caption{An example of Cyclic Dependency removal. Based on Lippert's example \cite[p. 128]{Lippert2006}.}
    \label{fig:cd-removal}
\end{figure}

As an example to better understand the reasoning behind this approach, let us take into consideration the case of a CD smell instance between two classes A and B (see Figure \ref{fig:cd-removal}). In order to remove it, the traditional way \cite[p. 128]{Lippert2006} is to split B in two (or more) segments $B_1$ and $B_2$ and separate dependencies in such a way that A depends on $B_1$ (or $A \rightarrow B_1$), and $B_2$ depends on A (or $B_2 \rightarrow A$).
This process implies that the developer must be familiar with all dependencies between A and B. For the sake of the example, let us assume that each line of code contains one dependency only. Then, we calculate, we count all the lines of code in A that use or create the dependency $A \rightarrow B$, and those for $B \rightarrow A$. Then we have $m(x) = m(A \rightarrow B) + m(B \rightarrow A) = 50 + 15 = 65$ LOC, which is the number of lines of code one needs to understand before deciding on how to split B and proceed with the refactoring. In Figure \ref{fig:cd-removal}, only $m(B \rightarrow A) = 15$ LOC were eventually moved to a new class, but the whole 65 lines of code were needed be understood before refactoring the 15 creating the dependency $B \rightarrow A$.

An architectural smell is comprised of several artefacts. Each artefact has a series of dependencies towards other artefacts, which we consider as edges (i.e. $A \rightarrow B$). We calculate 
\begin{equation}\label{eq:smell-extent-dependency}
    m(x) = \sum^{E_x} w(a \rightarrow b)
\end{equation}
where $$E_x = \{a \rightarrow b | a,b\textnormal{ are classes or packages affected by smell } x\}$$ and $w(a\rightarrow b)$ calculates the number of times artefact $a$ uses artefact $b$. 
By use we mean any time $a$ declares a variable of type $b$, invokes a method on an object of type $b$, accesses a field of an object of type $b$, or inherits from type $b$.
The way we calculate dependencies complies with the benchmark and guidelines provided by Pruijt et al. \cite{Pruijt2017}.

One can also see the $m$ function as a special, finer-grained case of the dependency edge weight function defined by Laval et al. \cite{Laval2012}, where instead of counting the import statements only, we count all the lines of code directly using such dependency.

An advantage of $m(x)$ is that it allows to handle the overlap between smells at a fine-grained level and avoid overestimation of the final effort calculated to remove all the smells (i.e. a single edge may be responsible for the creation of multiple smells). 
This is simply achieved using the following generalisation of Equation \ref{eq:smell-extent-dependency}:
\begin{equation}\label{eq:smell-extent-dependency-weight}
m(x) = \sum^{E_x} \frac{w(a\rightarrow b)}{o(a \rightarrow b)}
\end{equation}
where the contribution of each edge $a \rightarrow b$ is weighted by the number of smells that edge contributes creating, calculated by $o(a \rightarrow b)$.

Another advantage is that it allows to identify which edges yield the highest return on effort invested if removed, because one can target the edge with lowest use and highest number of smells passing through it.
Additionally, it allows to only include the edges that actually create the smell, for example, for UD smell, $m(x)$ may only include the edges that create dependencies towards less stable packages.

In \textsc{Arcan}, this feature is implemented by relying on Spoon \cite{Pawlak2015} to precisely calculate the lines of code generating a dependency (as defined by Pruijt et al. \cite{Pruijt2017}).

\paragraph{Size-based smells}
God Component is a smell that is detected based on the number of lines of code an artefact has (calculated summing up the LOC of the \emph{directly} contained files) and whether it exceeds a certain threshold.
The threshold is calculated using an adaptive statistical approach that takes into consideration  the number of LOC of the other packages in the system and in a benchmark of over 100 systems \cite{Fontana2015}.
The adaptive threshold is defined in such a way that it is always larger than the median lines of code of the packages/components in the system and benchmark.
Therefore, the goal of refactoring a God Component is to reduce the total number of lines of code in the system to be in line with the rest of the components in the system (i.e. get closer to the median of the system).
As mentioned earlier, the lines of code metric is a known predictor of complexity \cite{Lippert2006, Morasca2001, Kitchenham2004}, therefore to formalise this concept we define 
\begin{equation}\label{eq:smell-extent-gc-threshold}
    \delta(x) = LOC(x) - T_{median}   
\end{equation}
where $LOC(x)$ calculates the lines of code of in the artefact affected by the smell $x$, and $T_{median}$ is the median size of components in the system.

However, just the bare number of lines of code is not fully indicative of the effort.
The number of elements and the connection among those elements is a variable affecting the difficulty of performing such task. The more elements (and connections among them) there are in a component, the lower its Understandability \cite[p. 32]{Lippert2006} and the higher their coupling.
Therefore, we define the extent of a god component architectural smell as
\begin{equation}\label{eq:smell-extent-gc}
    m(x) = \delta(x) \cdot \sqrt{\frac{|E_x|}{2|V_x|}}
\end{equation}
where $|E_x| \ge 1$ and $|V_x| \ge 1$ are the number of edges and vertices respectively, contained in the subgraph created within the artefact affected by $x$.
The second term in Equation \ref{eq:smell-extent-gc} ensures that if there is loose coupling among the elements contained in the component affected by $x$, then the overall value is lower, because it is easier to identify what files to move to another component, or what files to split into multiple files before moving them. The square root is used to reduce the effect on the final result.
Indeed, early experimentation without the use of the second term resulted in over-estimations of the index in cases were the internal elements of a package were loosely coupled.

\subsubsection{Summary definition}
The $m$ function has a different definition based on the type of smell evaluated. To avoid misunderstandings, we formalise this in the present section by defining $m$ as follows:
\begin{equation}\label{eq:smell-extent-all}
    m(x) = \begin{cases}
        \sum^{E_x} \frac{w(a\rightarrow b)}{o(a \rightarrow b)} & \text{if $x$ is a CD, HL, or UD instance}\\
        \delta(x) \cdot \sqrt{\frac{|E_x|}{2|V_x|}} & \text{if $x$ is a GC instance}\\
    \end{cases} 
\end{equation}
where $x$ is an architectural smell instance, and the rest of the variables and functions are the same as defined in the previous sections.

\section{Case study design}\label{sec:study-design}
To evaluate the approach described in Section \ref{sec:approach}, we followed the guidelines proposed by Runeson et al. \cite{Runeson2009} to design an holistic multiple-case study.
Case studies are commonly used in software engineering research to study a phenomenon in its real-life context \cite{Runeson2009}.
We opted to perform a case study because it allows us to investigate the practical application of our approach in the context of both industrial and open source projects. 
In the next sections we elaborate on the study design.


\subsection{Goal and research questions}\label{sec:goal-and-rqs}
The objective of the case study is to evaluate the \emph{accuracy}, \emph{transparency}, and \emph{relevance} of our approach that estimates architectural technical debt principal using architectural smells.
Using the Goal-Question-Metric \cite{VanSolingen2002} formulation, the objective is stated as follows:
\begin{quote}
    \itshape
    \textbf{Analyse} the approach estimating architectural technical debt principal \textbf{for the purpose of} validating its application \textbf{with respect to} accuracy, transparency, and relevance of the estimation output \textbf{from the point of view of} software developers \textbf{in the context of} open source and industrial software systems.
\end{quote}
The goal can be further refined into the following two research questions, reflecting accuracy and relevance respectively:
\begin{itemize}
    \item[\textbf{RQ1}] Can the approach accurately and transparently rank architectural smells by their severity?
    \begin{itemize}
        \item[\textbf{RQ1.1}] How accurate is the ranking of AS by different ML models?
        \item[\textbf{RQ1.2}] How do smell characteristics impact the predictions of severity?
    \end{itemize}
\end{itemize}
Essentially, we are interested in the accuracy (whether the ranking by severity is close to the ground truth) and transparency (what information is used to rank an instance) of the output of the approach, i.e. the principal. Our approach uses two factors to estimate the principal of each smell instance: severity and extent. We do not need to validate the accuracy and transparency of the smell extent, as that can be measured directly on the source code generating the smell. 
Thus, RQ1.1 concerns the \emph{accuracy} of calculating smell severity, and particularly the accuracy of the machine learning model in ranking architectural smells by their severity.
We will assess the accuracy of the model using an evaluation metric specific to ranking tasks as described in Section \ref{sec:perf-metric}.
RQ1.2 focuses on measuring the \emph{transparency} of the machine learning model. Namely, it will explain how the model effectively makes predictions on new, unseen instances, thus allowing us to better understand which smell characteristics make a smell more severe than another.
This will also ensure that the model is not using undesired variables to predict the severity of a specific smell (e.g. the Shape characteristic is only used for CD instances, and should be be used to predict the severity of a HL).

\begin{itemize}
    \item[\textbf{RQ2}] Is the principal estimated by the approach relevant to software developers?    
    \begin{itemize}
        \item[\textbf{RQ2.1}] Does the estimated principal represent the effort necessary to refactor an architectural smell?
        \item[\textbf{RQ2.2}] Are the size and order of the estimations of individual smells meaningful in relation to each other? 
        \item[\textbf{RQ2.3}] What do software developers think about the proposed approach overall?
    \end{itemize} 
\end{itemize}
This research question assesses if the whole approach is relevant, in terms of providing an actionable output to developers (i.e. can they make decisions using the output provided by the approach?).
We answer this research question by answering the three sub-questions.
RQ2.1 focuses on how far the estimated principal correlates with the effort expected by the engineers to refactor a certain instance. 
This would allow engineers to \emph{reliably plan} the allocation of their resources (e.g. time) during the repayment phase.
RQ2.2 focuses on whether the approach allows comparisons between different smell instances (e.g. if this smell's estimated ATDI is $x$ than it makes sense for the other smell's ATDI to be $y$).
If indeed the magnitude and relative size of the estimations with respect to each other are meaningful, then the approach provides the means to make \emph{evidence-based} prioritisation decisions for resolving smells.
Finally, RQ2.3 aims at understanding the \emph{general opinion} of software developers towards architectural smell analysis and the estimations provided by ATDI.

\subsection{Overview of the case study}
\begin{figure*}
    \centering
    \includegraphics[width=0.8\textwidth]{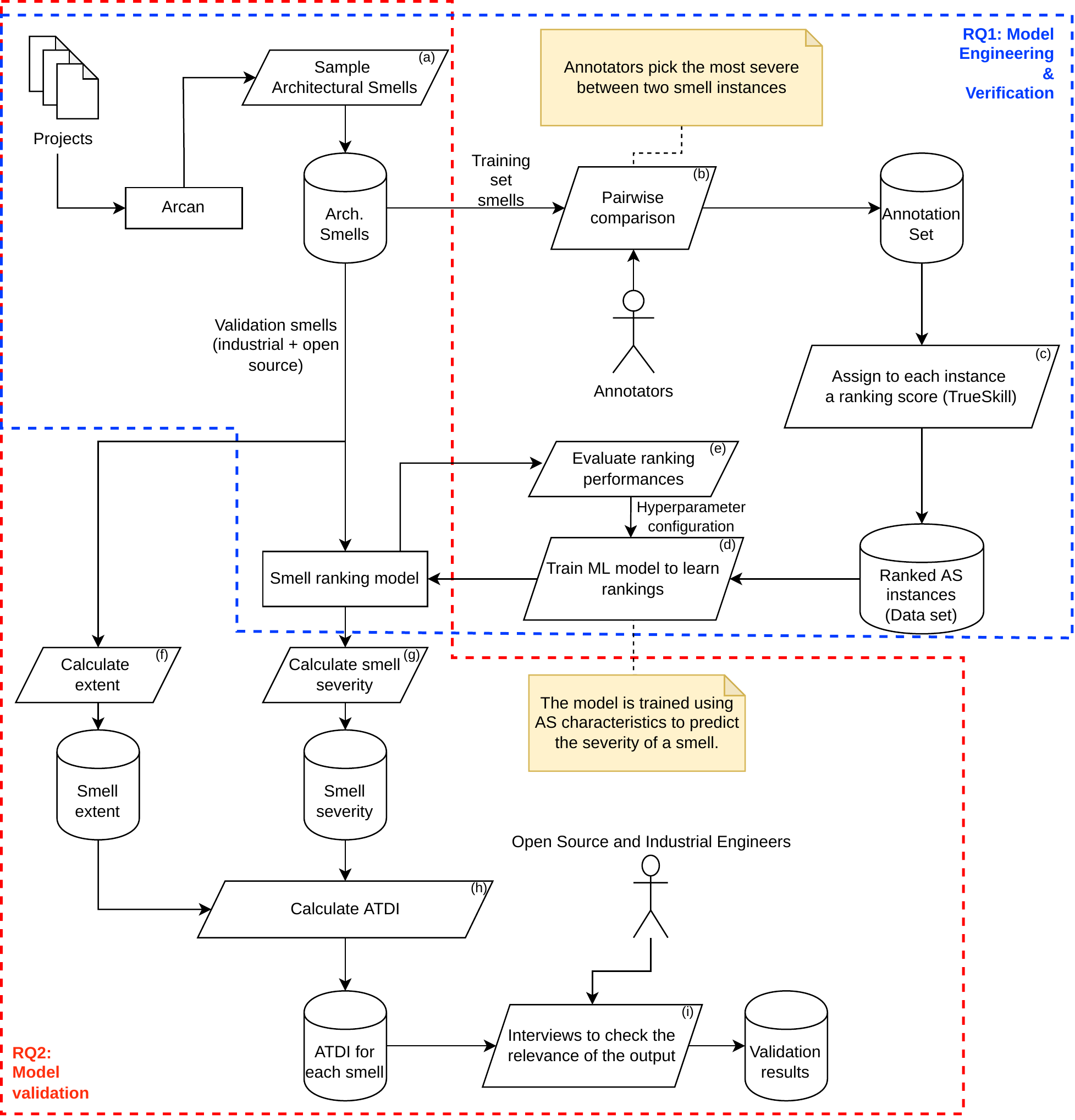}
    \caption{Detailed diagram of the model engineering and model evaluation phases.}\label{fig:study-overview}
\end{figure*}

The two research questions (RQ1 and RQ2) correspond, respectively, to two different phases of this study: \textbf{model engineering \& verification} and \textbf{model validation}.
Figure \ref{fig:study-overview} depicts a detailed overview of these two phases, while Section \ref{sec:rq1-methodology} and Section \ref{sec:rq2-methodology} respectively describe the two phases in detail.

The process for answering RQ1 can be summarised, using the steps from Figure \ref{fig:study-overview}, as follows: step \textit{(a)} concerns itself with the analysis of several software projects using \textsc{Arcan}; the results are used in steps \textit{(b)} and \textit{(c)} to create a dataset that ranks architectural smells by their severity; this dataset will be used in steps \textit{(d)} and \textit{(e)} to iteratively train and evaluate the performance of such model until its performance is satisfactory.
The output of these two steps is then used to answer RQ1.1 and RQ1.2 respectively.

Similarly, the process for answering RQ2 can be summarised as follows: the (partition of the) output of step \textit{(a)} that was not used to answer RQ1 is used in steps \textit{(f)} through \textit{(h)} to calculate ATDI for each smell; then, its output is used in step \textit{(i)} to answer RQ2 through a series of interviews with the practitioners that developed the systems containing the smells analysed.

The upcoming Sections \ref{sec:rq1-methodology} and \ref{sec:rq2-methodology} describe the research methodology for RQ1 and RQ2, respectively.

A replication package of this study is available online\cite{ReplicationPackage} and contains all the material used to design this case study.

\subsection{RQ1: Model engineering \& verification}\label{sec:rq1-methodology}
\subsubsection{Dataset creation}
\paragraph{Sampling the smells}
To answer RQ1 we trained a machine learning model to rank architectural smells by their severity.
The first step necessary to do so, as shown in Figure \ref{fig:study-overview}, step \emph{(a)} was to collect the data necessary to train the machine learning model, i.e. the architectural smells. This entailed choosing a set of projects to mine the smells from using \textsc{Arcan} (see Section \ref{sec:AS}). 
The selection criteria to choose the projects were the following:
\begin{enumerate}
    \item The projects must have more than 10.000 LOC; 
    \item The projects must have at least one instance of each architectural smell type;
    \item The annotators must be familiar with the architecture of the system they are annotating.
\end{enumerate}
These criteria ensured respectively that: 1) the projects selected were sufficiently big to contain enough architectural smells; 2) for each project there can be a comparison between all smell types; 3) the annotated smells affected parts of code that the annotators were familiar with, thus being able to provide a relevant annotation.
The projects selected by this process are shown in Table \ref{tab:smells-data-set}, along with the number of smells detected in each project.
Given that some architectural smell types (e.g. CD) exhibity higher occurrence rate than others, we randomly sampled at most 140 CD for each project and used stratified sampling to sample as many smells as possible from each project (which was dictated by the minimum occurring type of AS), as shown in Table \ref{tab:smells-data-set}. 
This avoided having an excessively predominant number of CD while also maintaining an organic distribution of the smell types in our sample.

\begin{table}[]
    \footnotesize
    \centering
    \caption{The open source projects (and Arcan) used for sampling smells and the number of smells compared as well as the number of comparisons.}
    \label{tab:smells-data-set}
    \begin{tabular}{@{}lm{1cm}m{1.1cm}m{1.1cm}m{1.1cm}m{1.1cm}@{}}
    \toprule
    \textbf{Project} & \textbf{LOC} &  \textbf{AS detected} & \textbf{AS sampled} & \textbf{Neighbours} & \textbf{Compar-isons} \\ \midrule
    Arcan & 30k & 192 & 55 & 14 & 23 \\
    AStracker & 10k & 77 & 28 & 7 & 7 \\
    Emma & 23k & 177 & 54 & 10 & 15 \\
    JMeter & 147k & 716 & 154 & 22 & 77 \\
    JUnit4 & 31k & 60 & 29 & 7 & 7 \\
    Spoon & 155k & 2859 & 155 & 22 & 77 \\
    Spring-boot & 366k & 133 & 92 & 15 & 35 \\
    Struts2 & 158k & 849 & 84 & 14 & 30 \\ \midrule
    \textbf{Total} & - & \textbf{5063} & \textbf{651} & \textbf{111} & \textbf{271} \\ \bottomrule
    \end{tabular}
\end{table}

\paragraph{Annotation set creation}
The smells sampled from the selected projects were then used to create an annotation set that contained, for each record, a pair of smells and an annotation denoting which one of them is the most severe one.
As one can see in Figure \ref{fig:study-overview}, step \emph{(b)}, annotations were manually created using \emph{pairwise comparison}, a process for annotating entities \cite{David1963} where an annotator is asked to compare two entities w.r.t. a certain quantitative property and provide a qualitative judgement on which one of the two entities is best.
The main reason for using pairwise comparisons over rating scales (e.g. Likert scale) is that it avoids several problems typical of rating scales. 
More specifically, rating scales are relative, which means that a value of 4 may not represent a similar quantity for two different individuals.
Also, the quantity represented for one individual may change during the questionnaire (e.g. after answering more questions) or if repeated in different days \cite{Perezortiz2017}, whereas, if two smells are compared twice and obtain discordant ratings, their final ranking will just depend more on the annotations where the two smells were compared with other smells.
These disadvantages make a rating scale, such as a Likert scale, a poor choice for this step.

The main drawback of pairwise comparison is the very large amount of comparisons necessary to achieve an order among the elements compared.
Pairwise comparisons necessitates $\binom{n}{2}$ comparisons. If we want to create a data set with $n = 500$ elements, then \emph{124.750} comparisons are necessary.
This number is infeasible for the purposes of our study; therefore, we adopted an array of techniques to reduce this number while at the same time increasing the number of elements in our data set:
\begin{enumerate}
    
    \item \emph{Active Sampling} is a technique that chooses the pairs to compare based on which one gives the most amount of information \cite{Mikhailiuk2020}. This technique is basically a compromise between number of comparisons and accuracy of the ranking with respect to the ground truth (i.e. the order obtained by doing $\binom{n}{2}$ comparisons). 
    The more comparisons are performed, the lower the error accumulated.
    Moreover, active sampling allows to reduce this error much faster than random selection of pairs to compare. 
    Several state-of-the-art techniques exist to perform this task, but ASAP \cite{Mikhailiuk2020} is the latest and fastest at the moment of writing.
    With this technique, we are guaranteed to reach at worse a 15\% error within $\frac{1}{3}\binom{n}{2}$ comparisons.
    
    This technique alone, however, is not sufficient to reduce the number of comparisons to a feasible amount.

    \item \emph{Initial ranking} gives an initial estimation of the final rank of the smell based on the architectural smells characteristics of each instance (i.e. the number of elements affected, number of dependencies, etc.). 
    The calculation of the initial ranking is based on previous work on architectural smell ranking \cite{Laval2012} and on smell characteristics \cite{Sas2019}.
    
    This allows to avoid comparisons of smells that are clearly at the two ends of the ranking range (e.g. a cycle of size 20 and a cycle of size 3).

    \item \emph{Neighbourhood Representative Sampling (NRS)} is based on the core concept behind the $k$-nearest neighbours ($k$-NN) algorithm, a classification and regression model widely used in machine learning \cite{Fix1989}: \emph{similar instances will probably have a similar classification}.
    This rationale can also be applied to the initial ranking, namely, \emph{similar instances will have a similar initial ranking}.
    Therefore, if we choose $k$ as the number of neighbourhoods and `appoint' one representative for each neighbourhood, we only have to compare $k$ elements rather than $n$.
    Obviously, the smaller the value of $k$, the more precise the final ranking.
    We selected $k$ with the following formula $k = \lfloor\log_2 n\rfloor$ for all $n \ge 5$, otherwise we used $k = 1$ (i.e. compared all smells).
    The number of representatives is reported in the third column of Table \ref{tab:smells-data-set}.

    \item \emph{Intra-project comparisons} entails comparing smells from the same project only. This is justified because during the analysis of a project, the ML model will rank smells from that project only.

\end{enumerate}
These four techniques are combined as follows: we first assign an initial ranking to each smell; then, we choose $k$ neighbourhoods and pick a smell that has the most similar initial ranking in that neighbourhood and designate it as its \emph{representative}; next, we perform pairwise comparisons among the representatives using active sampling until we obtain an order among the representatives; finally, the ranking is extended to the other smells in the neighbourhood.
This whole process is contained in Figure \ref{fig:study-overview} under the step \emph{(b)}, for the sake of simplicity.

The next question is \textbf{how to go from triplets} in the form of $\langle smell_1, smell_2, annotation \rangle$ (i.e. the output of the comparison process in step \emph{(b)}) \textbf{to a ranked order} among the elements compared -- which brings us to step \emph{(c)}.
There exist several algorithms that perform this task, but we opted for one of the most common solutions both in industry and academia, namely TrueSkill \cite{Herbrich2006}.
TrueSkill has been used extensively in information retrieval, learning-to-rank models, and even in software engineering to decide on how to assign tasks to components in simulation systems \cite{Wienss2013}, or to study the biases present in case studies analysing the language adoption of software developers \cite{Meyerovich2012}.
The TrueSkill algorithm seems the most pertinent for our purposes given its application in other software engineering research studies, as well as the wide availability of its implementation.

\paragraph{Data set creation and annotators agreement}
After completing step \emph{(c)} from Figure \ref{fig:study-overview}, we obtained a data set of 651 smell instances (see Table \ref{tab:smells-data-set}) that were ranked according to their severity, requiring 271 comparisons.
Comparisons required around 5 minutes each, and the whole process took 22 hours split among 3 annotators.
The annotation team was comprised of two Ph.D. students (including the first author) and a research assistant.
Inter-annotator agreement was measured using Fleiss's Kappa \cite{Fleiss1971}, obtaining a $.88$ score (considered `almost perfect agreement' \cite{Fleiss1971}).
Three test run rounds were necessary to achieve a score greater than $.8$ (i.e. greater than `moderate agreement' \cite{Fleiss1971}), with the first two rounds scoring $.29$ and $.43$. 
We ensured all three annotators used the same decision-making process to annotate the data by devising a set of rules, available in the replication package \cite{ReplicationPackage} along with the resulting annotations.
Those rules were based on the available literature \cite{Laval2012,Al-Mutawa2014}, our own experience on the subject, and the discussion among the annotators after each of the three test rounds.

\begin{figure}
    \centering
    \includegraphics[width=\linewidth]{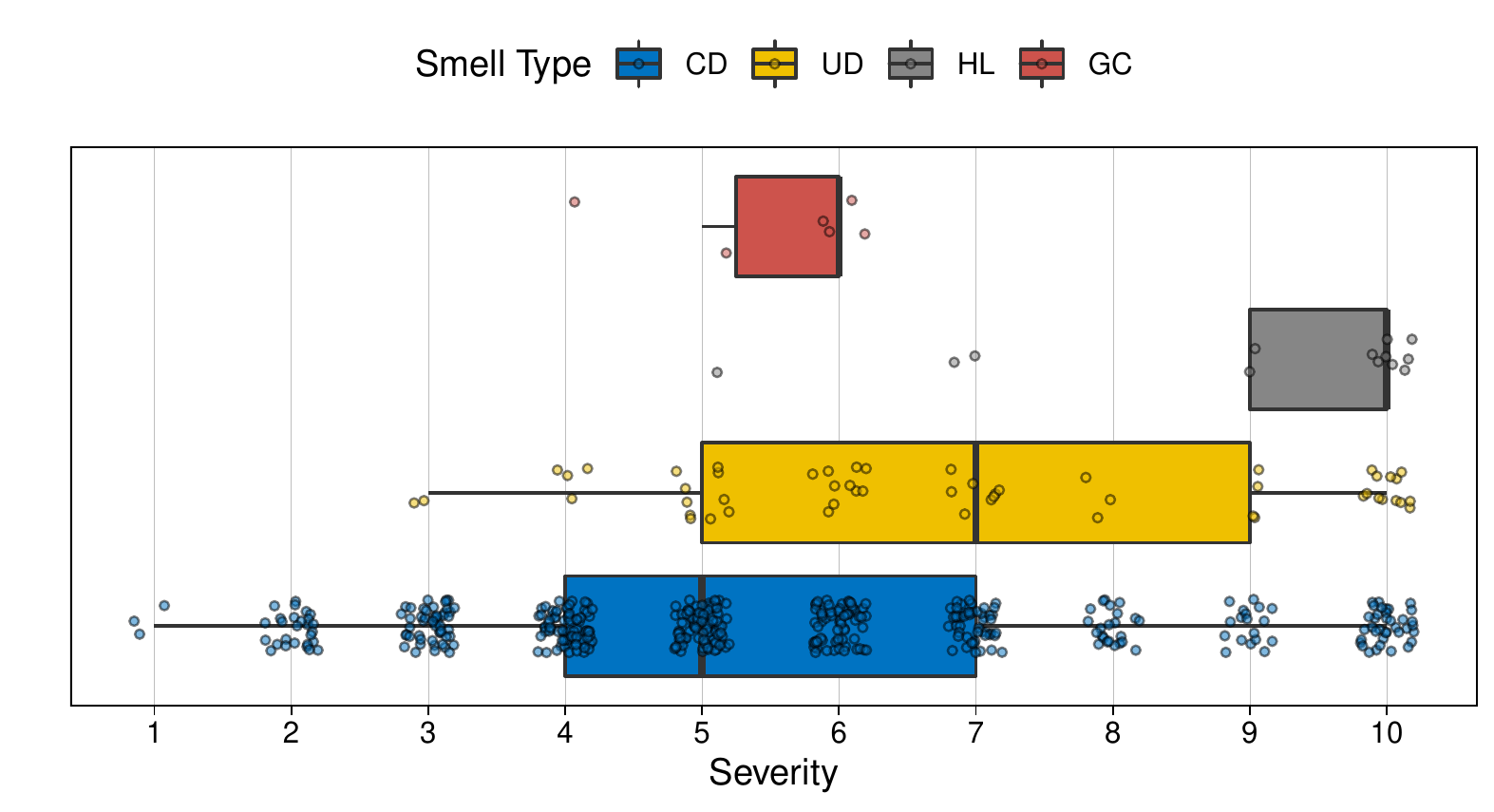}
    \caption{Distribution of (severity) labels obtained through our annotation process. Original data points are showed with slight jitter for better visualisation. Severity was rounded to 0 decimals in order to comply with LightGBM requirements.}\label{fig:dataset-label-distr}
\end{figure}

The distribution of the labels obtained through this process is depicted in Figure \ref{fig:dataset-label-distr}.
As it can be noted, CD smells are distributed almost perfectly across the domain of severity, whereas the other smells are skewed towards higher values.
This is because CD smells are much more easily detectable and there exist many more instances that pose little threat to the maintainability/evolvability of a system \cite{Al-Mutawa2014,Laval2012}.
This is, in contrast, rather unlikely for a GC or HL instance.
The distribution of the number of different types of smells is representative of the typical distribution found when analysing other software systems \cite{Sas2021}.

The training of the machine learning model was done using a 7-fold cross validation (step \emph{(d)}) and we evaluated ranking performance using normalised discounted cumulative gain (step \emph{(e)}).
This step (step \emph{(e)}) allows us to measure the accuracy of the ML model, i.e. to answer RQ1.
Further details on the training and performance obtained are reported in the next section.

\subsubsection{Training strategy \& evaluation metric}\label{sec:perf-metric}
To select the most suitable model for our task, we relied on the current state-of-the-art library for LTR tasks: LightGBM \cite{Ke2017}.
The training process used is $k$-fold cross-validation \cite{Stone1974}, a process where the data set is divided into $k$ equal partitions with one partition that acts as test set and the rest as training set; the process is then repeated until all $k$ partitions acted as test set.
The main advantage of using cross-validation over classic approaches such as plain train/test partitioning is the reduction of \emph{selection bias}, ensuring that the model performs similarly regardless of the seed used to partition the data set.

The metric that is most suitable to evaluate the performance of our model is Normalised Discounted Cumulative Gain ($NDCG$) \cite{Jarvelin2002}. 
$NDCG$ is the most common metric used in information retrieval to evaluate the efficiency of an algorithm to retrieve results in a certain order \cite{Wang2018}.
As an example of its use in software engineering studies, it was used to evaluate the relevance of algorithms retrieving architectural knowledge from StackOverflow \cite{Soliman2018}.

The goal of our task is to \textit{\textbf{minimise the number of times a severe smell is ranked below a less severe smell}}.
$NDCG$ matches perfectly our goal, as it penalises smells appearing lower than less severe smells in a ranked result list.

The formula of NDCG is as follows
$$NDCG = \frac{DCG}{IDCG} = \frac{1}{IDCG}\sum_{i=1}^n\frac{\ell_i}{\log_2(i+1)}$$
where $\ell_i$ is the severity label of the smell at position $i$, $DCG$ is the discounted cumulative gain, and $IDCG$ is the $DCG$ calculated on the sequence of retrieved elements in the ideal order (i.e. we sort results by $\ell_i$ such that smells with higher values of $\ell_i$ appear first).

The $NDCG$ metric (unlike $DCG$) is defined in the interval $[0, 1]$, with higher values meaning better performance/ranking of results.
In most scenarios, $NDCG$ is calculated only for the first $n$ elements of the test set, denoted as $NDCG@n$.
By combining multiple measures of $NDCG@n$ for different values of $n$, one can gauge the performance on incremental sub-lists of the result.
In other words, $n$ restricts the focus on the performance obtained by classifying the top $n$ most severe smells in the test set.

\subsection{RQ2: Model validation}\label{sec:rq2-methodology}
Figure \ref{fig:study-overview} depicts the process used for the validation of the model (red frame).
In particular, we detect architectural smells in open source and industrial systems, use the machine learning model developed in RQ1, calculate $ATDI$ through calculating extent and severity, and then collect the opinions of software practitioners about the output. 
The opinions are solicited through interviews, which, as a direct data collection technique, allows researchers to control exactly what data is collected, how it is collected, and in what form it is collected \cite{Runeson2009,Lethbridge2005}. 

\subsubsection{Cases, subjects and units of analysis}
\begin{figure}
    \centering
    \includegraphics[width=\linewidth]{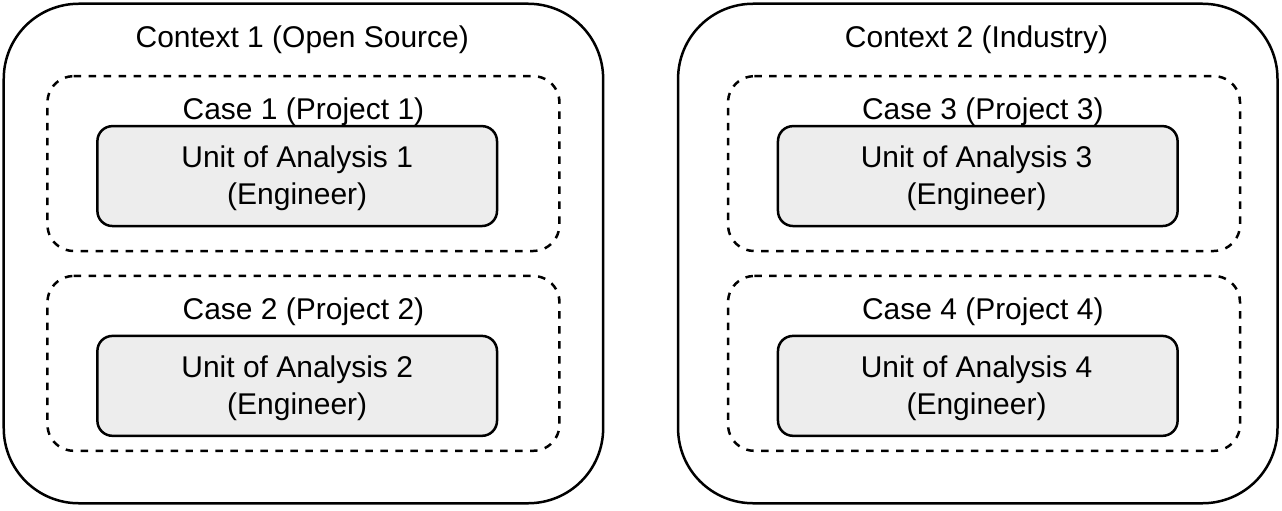}
    \caption{Mapping of cases and units of analysis for RQ2; based on Figure 3.1 by Runeson et al. \cite{Runeson2009}.}\label{fig:case-study-design-rq2}
\end{figure}

The cases of our study are the projects analysed whereas the context is either open source or industry; Figure \ref{fig:case-study-design-rq2} illustrates as an example, two cases from each context, from a total of sixteen cases. 
Finally, the units of analysis correspond to the software practitioners interviewed. Since each case contains a single unit of analysis, the design of the case study is multiple and holistic (see Runeson et al. \cite{Runeson2009}).

Tables \ref{tab:open-source-participants} and \ref{tab:industrial-participants} list the participants of the interviews, alongside their respective background information; the sample contains 9 engineers from open source projects and 7 from industrial projects.
We opted to interview one engineer per project (both for Open Source and  industrial projects) in order to maximise the variance of information obtained and avoid overlaps, thus extending external validity.
Note that most of the participants from open source projects are also employed in industry.

The open source participants were selected through the following process:
\begin{enumerate}
    \item We first collected a list of open source projects featured in other recent studies on Technical Debt that involved interviews and/or surveys \cite{Tan2021,Maldonado2017,Zampetti2021}. This ensured that the candidate projects contained technical debt and resulted in selecting 21 open source projects (listed in Figure \ref{fig:atdi-distribution});
    \item We listed the most active contributors from the repositories of such projects (top 10\% of number of commits in the last year). We selected the most active contributors to ensure that they had deep understanding of the system (or of a specific part of it) and that they were up-to-date with the latest code. 
    This resulted in over 260 contacts, and after removing bots and invalid emails we ended up with 230 contacts;
    \item We sent out 230 invitation emails and received 37 responses, of which 11 of them contained a positive response and eventually 9 resulted in an interview.
\end{enumerate}

To select the industrial participants, we used purposeful sampling \cite{Palinkas2015}. Specifically, we got in touch with two companies from our professional network and asked them whether they were willing to participate in the study. 
The two companies are both small and medium-sized enterprises\footnote{See \url{https://ec.europa.eu/growth/smes/sme-definition_en}.} that operate in the IoT and Enterprise Application domains, respectively.
Next, we asked them to provide us with (1) a list of Java projects that had at least 10.000 lines of code, and (2) a list of engineers working on these projects that were willing to take part in the interviews.

Overall, the sample is comprised of 16 engineers (and their respective projects), characterised by a wide variety in total number of years of experience and technological background (e.g. distributed systems, testing, security, etc.).
Of course, no sample is perfect, and we elaborate on the threats to external validity entailed by the composition of our sample in Section \ref{sec:threats-to-validity}.

\begin{table}[]
    \centering
    \footnotesize
    \caption{List of participants from the open source projects. Note that the `Role in project' column was \textbf{shuffled} to protect the anonymity of the participants. For example, P1 is not an idependent contractor, but one of the other participants is. Abbreviations: \textbf{Partic.}: participant; \textbf{OS}: open source; \textbf{IN}: industry; \textbf{Exp.}: experience; \textbf{PMC}: Project Management Committee; \textbf{MC}: Main Contributor.}
    \label{tab:open-source-participants}
    \begin{tabular}{@{}clm{2cm}l@{}}
    \toprule
    \textbf{Partic.} & \textbf{Project} & \textbf{Exp. in} \textbf{OS/IN (Years)} & \textbf{Role in project} \\ \midrule
    P1 & Hadoop & 12 / 8 & \multirow{9}{*}{\begin{tabular}[c]{@{}l@{}}Independent Contractor\\ Team lead and MC\\ PMC member \& Contributor \\ Security Engineer \\ PMC member \\ PMC member \& contributor\\ Lead maintainer\\ Contributor\\ Project lead and MC\end{tabular}} \\
    P2 & DBeaver & 5 / 18 &  \\
    P3 & JUnit5 & 12 / 14 &  \\
    P4 & RxJava & 10 / 15 &  \\
    P5 & Jenkins & 18 / 11 &  \\
    P6 & Hibernate & 20 / 20 &  \\
    P7 & Cassandra & 8 / 24 &  \\
    P8 & Camel & 18 / 20 &  \\ 
    P9 & HBase & 16 / 20 &  \\ 
    \midrule
    \multicolumn{2}{r}{Average} & 13.1 / 16.6 &  \\ \bottomrule
    \end{tabular}
\end{table}

\begin{table}[]
    \centering
    \footnotesize
    \caption{List of participants from the industrial projects. Abbreviations: \textbf{mgmnt.}: management; \textbf{Partic.}: participant; \textbf{Exp.}: experience.}
    \label{tab:industrial-participants}
    \begin{tabular}{@{}ccm{2cm}m{0.5cm}l@{}}
    \toprule
    \textbf{Partic.} & \textbf{Company} & \textbf{Project} & \textbf{Exp. (Years)} & \multicolumn{1}{c}{\textbf{Role}} \\ \midrule
    P10 & C1 & IoT Framework & 3 & Developer \\ 
    P11 & C1 & Document mgmnt. system & 15 & Senior developer \\ 
    P12 & C2 & Project mgmnt. tool & 22 & Product manager\\ 
    P13 & C2 & Rent mgmnt. API service & 8 & Senior developer \\ 
    P14 & C2 & Parking occupancy meter & 6 & Full-Stack developer \\ 
    P15 & C2 & Financial assets mgmnt. & 6 & Senior developer \\ 
    P16 & C2 & Subscription mgmnt. & 3 & Developer \\ \midrule 
    \multicolumn{3}{r}{Average} & 9 &  \\ \bottomrule
    \end{tabular}
\end{table}

\subsubsection{Data collection}
Interviews were held following the guidelines mentioned by Runeson et al. \cite{Runeson2009}.
Overall, the data collected for each participant are the following: (1) background information regarding their expertise; (2) whether they find the estimated principal to be representative of the required refactoring effort (RQ2.1); (3) whether they think the \emph{order and proportions} of the principal estimations were consistent among the instances presented (RQ2.2); (4) the rationale behind their answers on points (2) and (3);  and (5) their feedback on the whole analysis (RQ2.3).

The interviews lasted 30-35 minutes and were semi-structured in their format, meaning that the interviewer could deviate from the original list of questions if a certain answer given by the participant was interesting to explore in more depth. The replication package contains the interview invitation and the questionnaire with the list of questions \cite{ReplicationPackage}.
Each interview invitation contained (1) a one-pager with the definitions of the smell types discussed in the interviews; and (2) a letter informing the participant of the confidentiality of the interview as well as their right to not answer any question they do not wish to answer \cite{Runeson2009}.
Before the interview started, both aforementioned points were reiterated to the participants to ensure that they were familiar with the technical concepts discussed during the interview and that they agreed with the terms of the interview.
 
During the interviews, we showed the participants one instance of each architectural smell type. If one type was not detected in the particular system, we replaced it with an instance of a type already included, so as to ensure we collect the same amount of data from every engineer. 
Smells were chosen from parts of the system that the participants indicated to be most familiar with.
The smells were visualised graphically as a network where nodes corresponded to classes and packages, and edges corresponded to the dependencies among them.
Each smell was accompanied by a number representing the \textbf{effort necessary to refactor} that smell (i.e. the ATDI) and each participant was instructed that ATDI was an estimation of the effort to refactor.
Next, each participant was asked whether they agree with the information presented for each instance while also keeping in mind the estimations provided for the other instances.
This ensured that their answers were consistent among different smell instances.
Finally, each participant was asked to explain their answer and particularly their rationale.
This process allowed us to minimise the amount of explanation provided to the participants (reducing the risk of confusion and bias).

\subsubsection{Data analysis}
To analyse the data collected through the interviews, we adopted the Constant Comparative Method (CCM) \cite{Glaser2017,Boeije2002}, which is part of Grounded Theory \cite{Glaser1968}. Grounded Theory (GT) is one of the most important methods in the field of qualitative data analysis. 
It has been used extensively within both social sciences and software engineering and provides a structured approach to process and analyse the data collected from multiple sources.
GT increases the theoretical sensitivity of the researcher as the data analysis progresses and eventually allows to formulate hypotheses and theory \cite{Glaser1968}.

The CCM is an inductive data coding and categorisation process that allows a unit of data (e.g., interview transcript, observation, document) to be analysed and broken into codes based on emerging themes and concepts; these are then organised into categories that reflect an analytic understanding of the coded entities \cite{Mathison2005}.

\begin{figure}
    \centering
    \includegraphics[width=\linewidth]{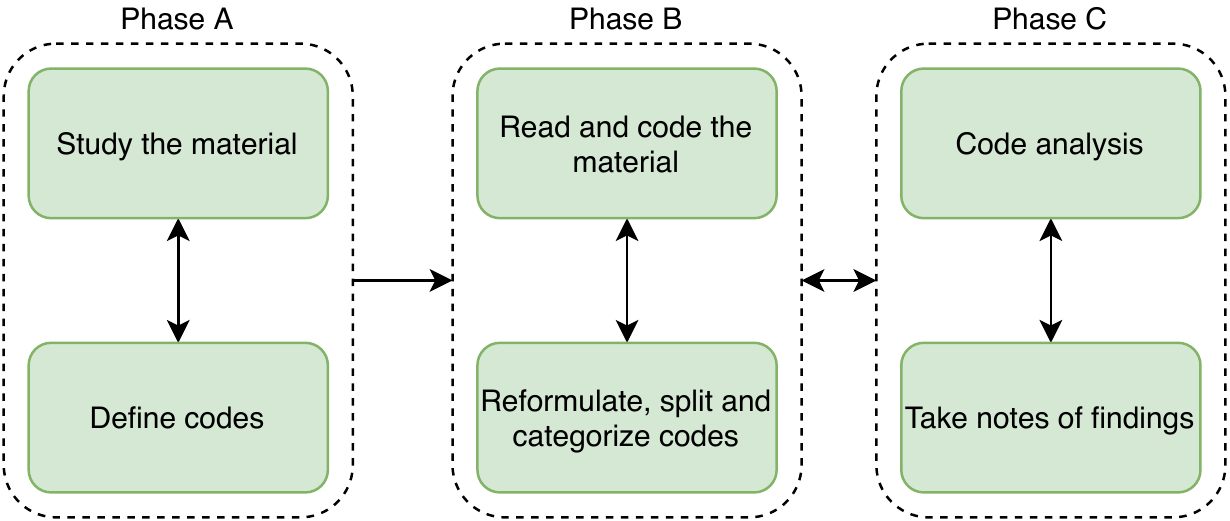}
    \caption{The qualitative data analysis process.}
    \label{fig:qualitative-analysis}
\end{figure}

The qualitative data analysis requires interviews to be transcribed before any of the techniques mentioned above could be applied.
Transcriptions were done as soon as batches of 2-3 interviews were completed, whereas data analysis was done iteratively.
Each iteration of the data analysis process is presented in Figure \ref{fig:qualitative-analysis} and is comprised of 3 phases. During the first phase (Phase A), the collected material (i.e. the initial interview transcripts) was studied and a code map was created to organise the codes used to tag the data.
After completing this phase, the coding process started (Phase B), which also involved updating and re-organising the codes based on the new understanding of the data.
Gradually, more interviews were recorded, transcribed, and coded and notes were taken with the aid of the coded data (Phase C).
In total, three iterations of data analysis were done (i.e. three times the whole process from Figure \ref{fig:qualitative-analysis}): the first for the interviews with open source engineers, the second with the industrial engineers, and the third to ensure that the codes added along the way were present in all the data.
The whole process was performed by the first author of the paper, while the second author reviewed the codes and coding schemes as they were developed to reduce the risk of biases (e.g. confirmation and information bias).
To automate the data analysis as much as possible, we relied on Atlas.ti\footnote{See \url{https://atlasti.com/}.}, a dedicated qualitative data analysis tool.

\section{Descriptive statistics of ATDI}\label{sec:descriptive-statistics}
Before presenting the results of the two research questions, we briefly present some descriptive statistics about ATDI and derive some observations.
These should provide more context on the results of both RQ1 and RQ2 and allow us to understand the statistical nature of the estimations provided by the approach.
Additionally, we also briefly compare the values we obtain by using the architectural smells detected by a tool different than \textsc{Arcan}.

\begin{figure*}
    \centering
    \includegraphics[width=\textwidth]{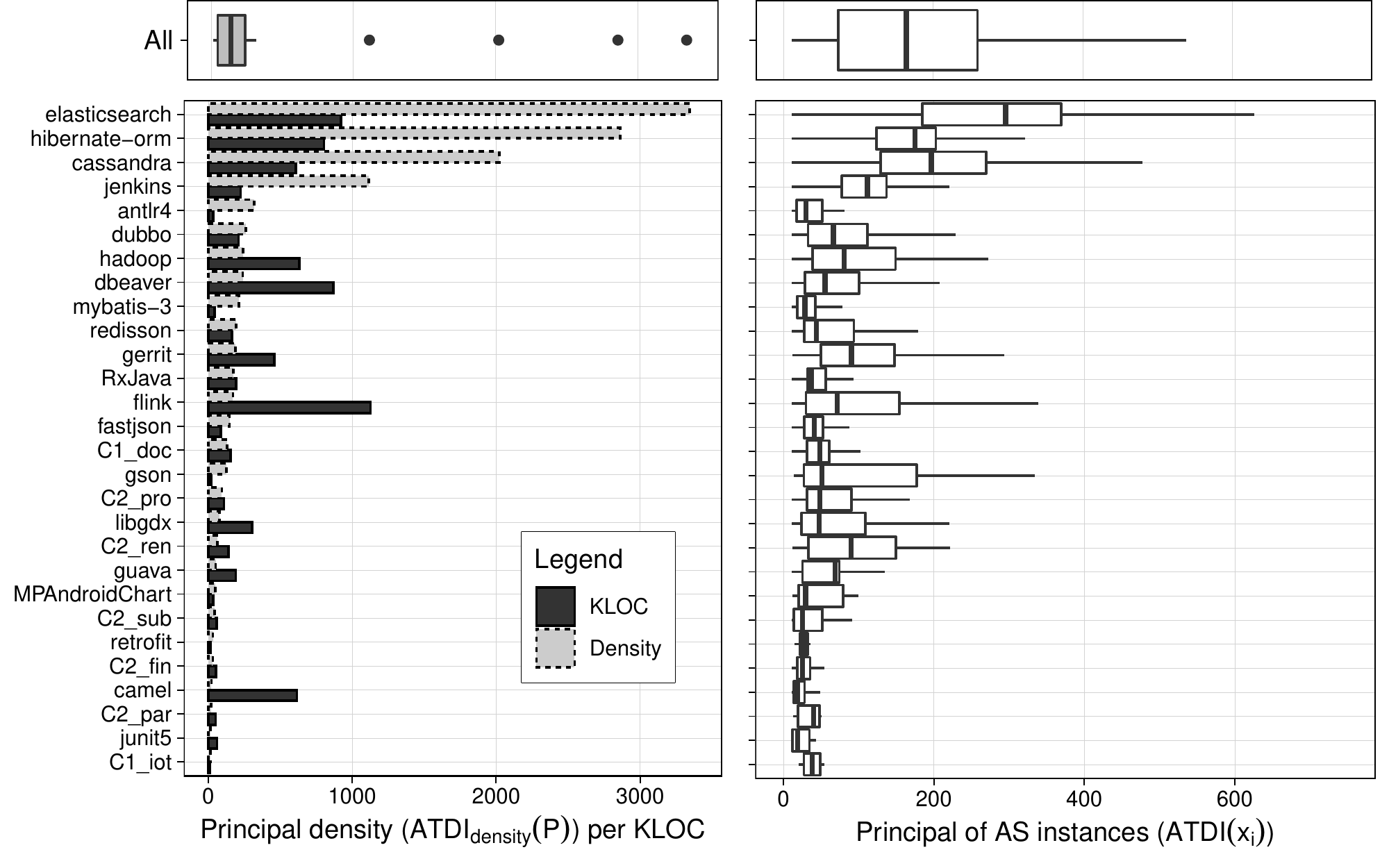}
    \caption{On the left, the total amount of principal (ATDI) per 1.000 lines of code (KLOC) for each project (calculated using Equation \ref{eq:atdi-normalised}) compared with the number of KLOC. 
    On the right, boxplots depicting the distribution of the principal (ATDI) calculated for each AS instance (outliers not visualised). }\label{fig:atdi-distribution}
\end{figure*}

These statistics concern the same 21 projects from which we collected the names of the open source participants for RQ2 as well as the 7 industrial projects; in total, ATDI was calculated for more than 41.000 smell instances of these 28 projects.
Figure \ref{fig:atdi-distribution} shows both the values of ATDI for each architectural smell instance and the value of ATDI density for the 28 projects considered.
The left-hand side plot depicts the total ATDI density for all projects, ordered from the most ATDI-dense project to the least.
The right-hand side plot depicts the distribution of ATDI for each AS instance in the 28 projects.
From the statistical analysis of the data depicted in Figure \ref{fig:atdi-distribution}, we note the following:
\begin{enumerate}
    \item the highest density project is ElasticSearch with 3345.8 ATDI for each KLOC, despite being the second largest system analysed;
    \item the lowest density project is JUnit5, with 16.2 ATDI for each KLOC;
    \item an overall lower ATDI density in a project does not always imply smells with lower individual ATDI. In particular, projects with lower ATDI density than \emph{Antlr4} (i.e. below it in Figure \ref{fig:atdi-distribution}), show a large variance in the ATDI of the individual instances;
    \item 50\% of AS instances have $ATDI \le 161$, and 33\% of instances have $ATDI \le 100$;
    \item there are only 37 smells with an $ATDI \ge 750$ (less than 0.001\% of all smells analysed);
    \item the maximum ATDI is 8505 by a HL smell in Jenkins;
    \item the minimum ATDI is 11, by three CDs with low severity in Camel, Cassandra and Dubbo respectively;
\end{enumerate}

\begin{figure}
    \centering
    \includegraphics[width=.9\linewidth,trim=2cm 0cm 0cm 0cm]{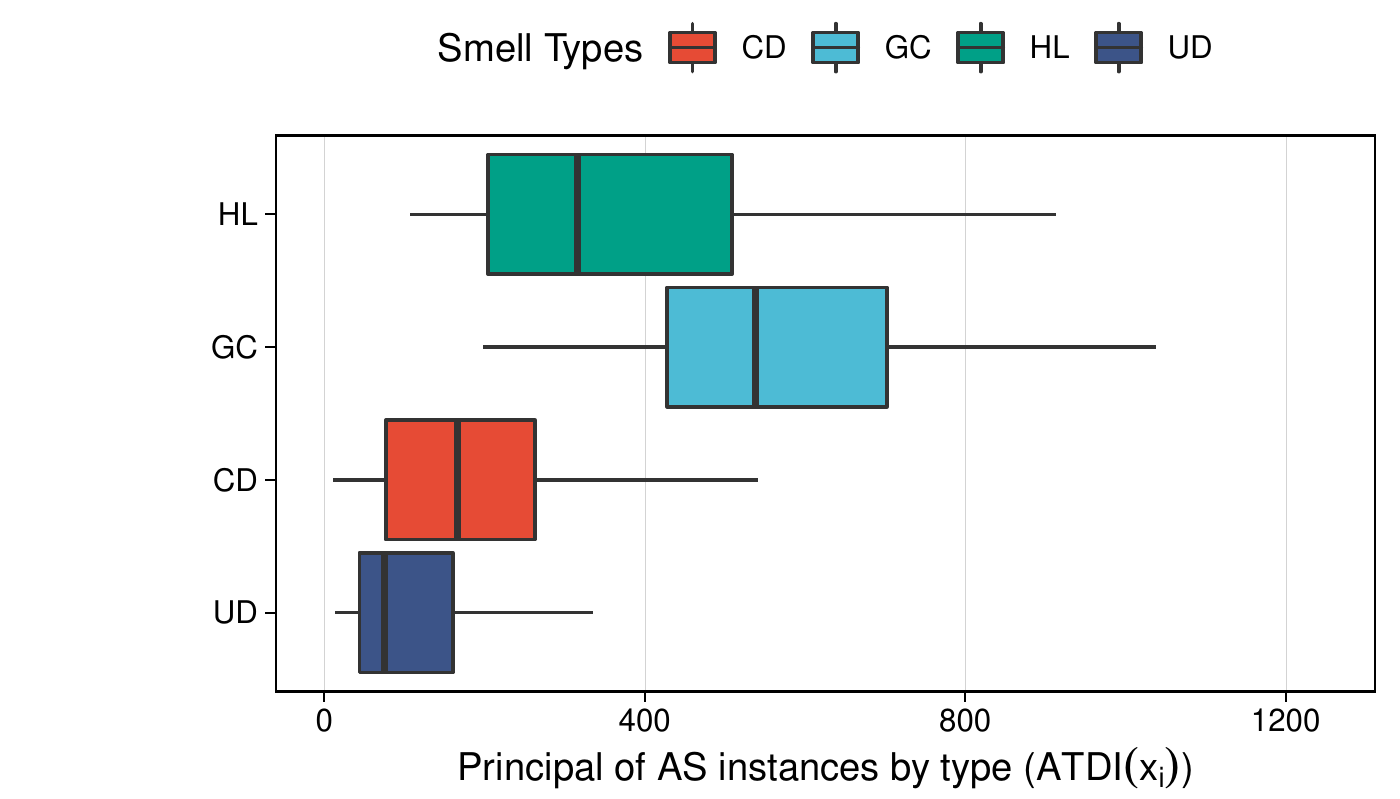}
    \caption{Boxplots showing the distribution of ATDI for different types of AS (outliers not visualised).}\label{fig:atdi-distribution-types}
\end{figure}

Figure \ref{fig:atdi-distribution-types} depicts the distribution of ATDI for different types of AS.
There is a clear difference between the four types of AS.
GC instances are the ones with the highest ATDI principal \emph{on average}, followed by HL, CD and lastly UD.

Finally, we briefly look at how different detection tools may impact the amount of TD calculated by ATDI. 
To do so, we analysed 10 of the systems from Figure \ref{fig:atdi-distribution} with Designite \cite{Sharma2016}, and used the smells detected as input to ATDI.
The results show that using different tools result in different values of ATDI for the same systems, namely the two tools result in two different distributions of ATDI (according with Wilcoxon signed ranks test). 
Despite this difference, the two samples (ATDI calculated with Designite and ATDI calculated with \textsc{Arcan}) are strongly correlated with a Spearman correlation coefficient $\rho=0.79$.
This means using different tools to calculate ATDI is \textbf{unlikely to significantly affect the conclusions} obtained.
Further details on the methodology and results of this analysis are available in the replication package \cite{ReplicationPackage}.

\section{RQ1 results}\label{sec:rq1-results}
\subsection{RQ1.1: ML model accuracy}
Table \ref{tab:algorithm-performances} summarises the $NDCG@n$ values obtained through cross-validation on our data set by different models.
We recall from Section \ref{sec:perf-metric} that $NDCG@n$ provides higher values when the model consistently ranks severe instances above less severe ones.
Note that we opted for 7-fold cross-validation\footnote{Folds are sampled using stratified sampling on severity, which is a typical practice in Machine Learning.} over the typical 10-fold because in our case it increases the size of the test set significantly (from 65 to 93) while it also reduces overfitting (i.e. we obtain much lower variance with $k = 7$).
The results show that the best-performing algorithm is `rank\_xendcg', i.e. Cross-Entropy NDCG Loss for learning-to-rank \cite{Bruch2021}, one of the most recent and best-performing LTR algorithms.
We refer the reader to the official documentation of LightGBM for details on the other algorithms\footnote{Visit \url{https://lightgbm.readthedocs.io/en/latest/Parameters.html\#objective}.}.

Overall, `rank\_xendcg' performs very well for all values of $n$.
However, the most severe smell is not always the very first smell in the list, but it does appear very close to the top in several occasions given the score obtained for $NDCG@1 = .99$.
For values of $n > 1$, `rank\_xendcg' settles around $.90$, meaning that most instances are ranked \emph{close} to their true rank, but not all of them.
When considering the order obtained on the full size of the training sets ($n = 93$), the performance reaches $.97$.
This means that the most-severe instance is \emph{almost perfectly ranked}, the mid-severity instances are \emph{appropriately ranked} but not quite perfect, while the low-severity instances are \emph{almost perfectly ranked}.

\begin{figure}
    \centering
    \begin{subfigure}[b]{0.7\linewidth}
        \centering
        \includegraphics[width=\linewidth]{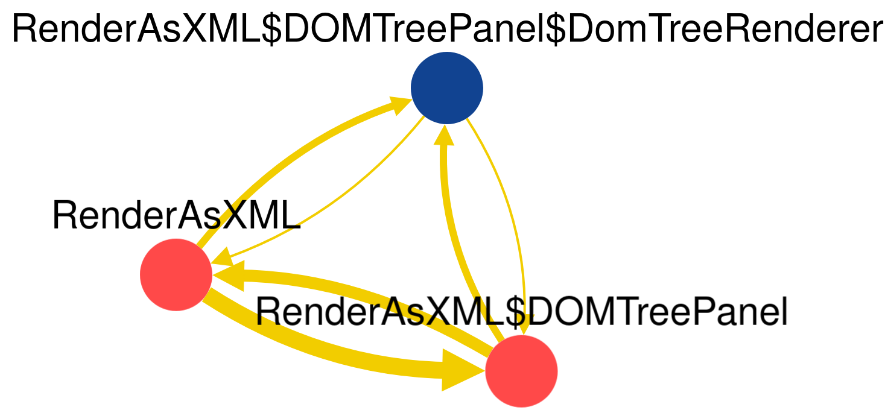}
        \caption{CD smell; Predicted: $1.84$; Actual: $1$.}
        \label{fig:severity-low}
    \end{subfigure}
    \hfill
    \begin{subfigure}[b]{.8\linewidth}
        \centering
        \includegraphics[width=\linewidth]{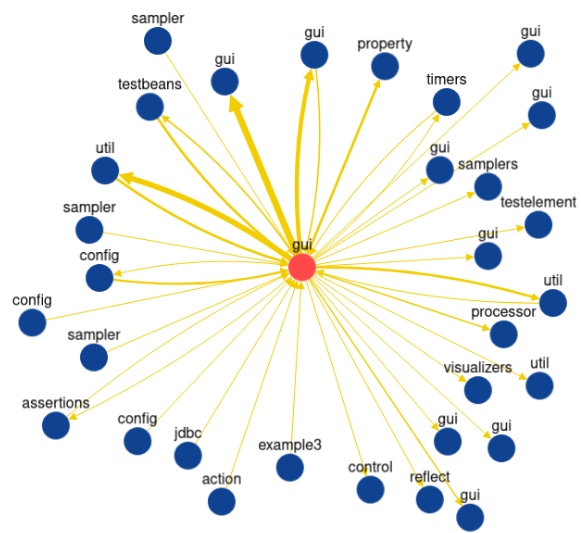}
        \caption{HL smell; Predicted: $9.11$; Actual: $10$.}
        \label{fig:severity-high}
    \end{subfigure}
    \\
    \begin{subfigure}[b]{.7\linewidth}
        \centering
        \includegraphics[width=0.7\linewidth]{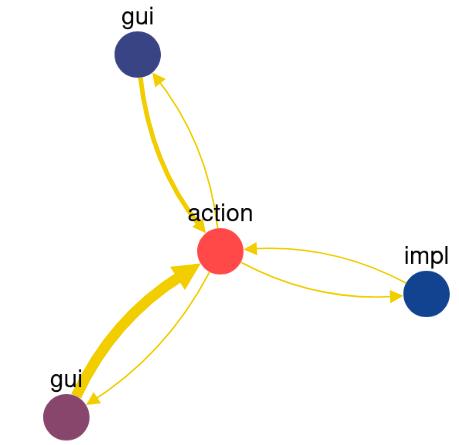}
        \caption{CD smell; Predicted: $5.28$; Actual: $5$.}
        \label{fig:severity-medium-1}
    \end{subfigure}
    \hfill
    \begin{subfigure}[b]{.7\linewidth}
        \centering
        \includegraphics[width=0.6\linewidth]{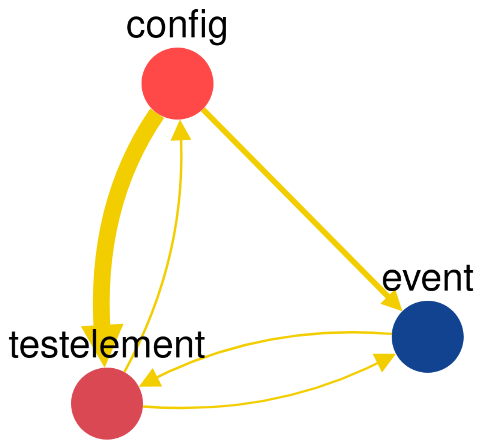}
        \caption{CD smell; Predicted: $5.46$; Actual: $5$.}
        \label{fig:severity-medium-2}
    \end{subfigure}
       \caption{Example of architectural smells from our data set with their predicted and actual severity. Smells are all from the JMeter project. The width of the edges reflects the weight of the dependency, while the colour of nodes reflects the number of weighted incident edges (red means higher values; blue lower).}
       \label{fig:severity-all}
\end{figure}

To make these results clearer, we give four examples of smells from our data set in Figure \ref{fig:severity-all}; their actual severity is obtained through the process described in Section \ref{sec:rq1-methodology}, while the predicted one by the ML model.
Figure \ref{fig:severity-low} depicts a CD smell with very low severity that affects one class and two of its internal classes.
Typically, this type of cycle is intentional, and given that the three classes are always expected to be reused together, this cycle does not pose any threat to maintainability, so it was labelled with the minimum severity of 1. 
The value predicted by the model was $1.84$, which is almost double the actual value, but it is still rather close.

Figure \ref{fig:severity-high} shows a rather severe HL instance affecting the main \texttt{gui} package in the system\footnote{Note that there are several packages called \texttt{gui} in JMeter.} and involving 31 other packages.
Given that \texttt{gui} is aggregating a lot of functionality (when in theory it should only be responsible for the user interface) it was annotated with a severity of 10. 
The model's prediction was a bit lower at $9.11$.

Figures \ref{fig:severity-medium-1} and \ref{fig:severity-medium-2} depict two smells of medium severity, both are CD smells affecting 4 and 3 packages, respectively.
Both smells were labelled as medium severity of 5, because they affect packages of the system, are tightly coupled, but are not too big in number of elements affected.
The predictions provided by our model for both smells were relatively close to the actual values.

To summarise, the goal of RQ1 was to check whether a ML model can accurately rank AS by their severity. 
This is indeed the case and we were able to achieve a score of $0.97$ for $NDCG@93$, which is considered a very high score. However, there is one caveat. 
When considering the accuracy of estimating severity for  single instances  (rather than the overall rank) the model is less accurate, as shown by the examples in Figure \ref{fig:severity-all}.
Namely, the model is clearly able to predict the ranking of the smells correctly, but the accuracy of the prediction is not perfect (Figure \ref*{fig:severity-low}).
Nevertheless, this is both expected and acceptable, as the goal for RQ1, was to optimise for the \emph{global ranking} of instances rather than the individual prediction. Indeed, this is also what the ML model is optimising for.

\begin{table}[]
    \centering
    \caption{Performance of different algorithms for different values of $NDCG$@$n$ using 7-fold cross-validation and the standard deviation over the folds. Bold values represent the maximum in the row.}\label{tab:algorithm-performances}
    \footnotesize
    \begin{tabular}{ccccm{.85cm}m{.85cm}}\toprule
    \multicolumn{1}{c}{\multirow{3}{.5cm}{$NDCG$\\@$n$}} & \multicolumn{5}{c}{\textbf{Algorithms}} \\
    \multicolumn{1}{c}{} & \textbf{mse} & \textbf{multiclass} & \textbf{multiova} & \textbf{rank\_\newline xendcg} & \textbf{lambda-rank} \\ \midrule
    1   &  .91$\pm$.00 &  .77$\pm$.04 &  .94$\pm$.01 &  \textbf{.99$\pm$.00} &  .99$\pm$.00 \\
    10  &  .87$\pm$.00 &  .86$\pm$.01 &  .88$\pm$.01 &  \textbf{.90$\pm$.00} &  .82$\pm$.00 \\
    25  &  .89$\pm$.00 &  .87$\pm$.00 &  .87$\pm$.00 &  \textbf{.90$\pm$.00} &  .87$\pm$.00 \\
    50  &  \textbf{.93$\pm$.00} &  .91$\pm$.00 &  .92$\pm$.00 &  .92$\pm$.00 &  .90$\pm$.00 \\
    93* &  .96$\pm$.00 &  .95$\pm$.00 &  .96$\pm$.00 &  \textbf{.97$\pm$.00} &  .95$\pm$.00 \\
    \bottomrule
    \end{tabular}\\
    * size of the test sets for $k=7$
\end{table}

\subsection{RQ1.2: The contribution of smell characteristics to predictions}\label{sec:explain-predictions}
Our model is able to predict quite accurately the severity of an architectural smell instance; however, it is also important to understand what smell characteristics are used, and how they impact the prediction. Namely, we want to improve the \emph{transparency} of the model by studying how it performs the predictions.
To do so, we used an approach called SHAP (SHapley Additive exPlanations). SHAP uses game theory to link the input of a model (i.e. the smell characteristics) to its output (i.e. the severity) \cite{Strumbelj2014} and explore the correlation visually.

\begin{figure}
    \centering
    \includegraphics[clip, trim=1.1cm 2cm -0.5cm 0cm, width=\linewidth]{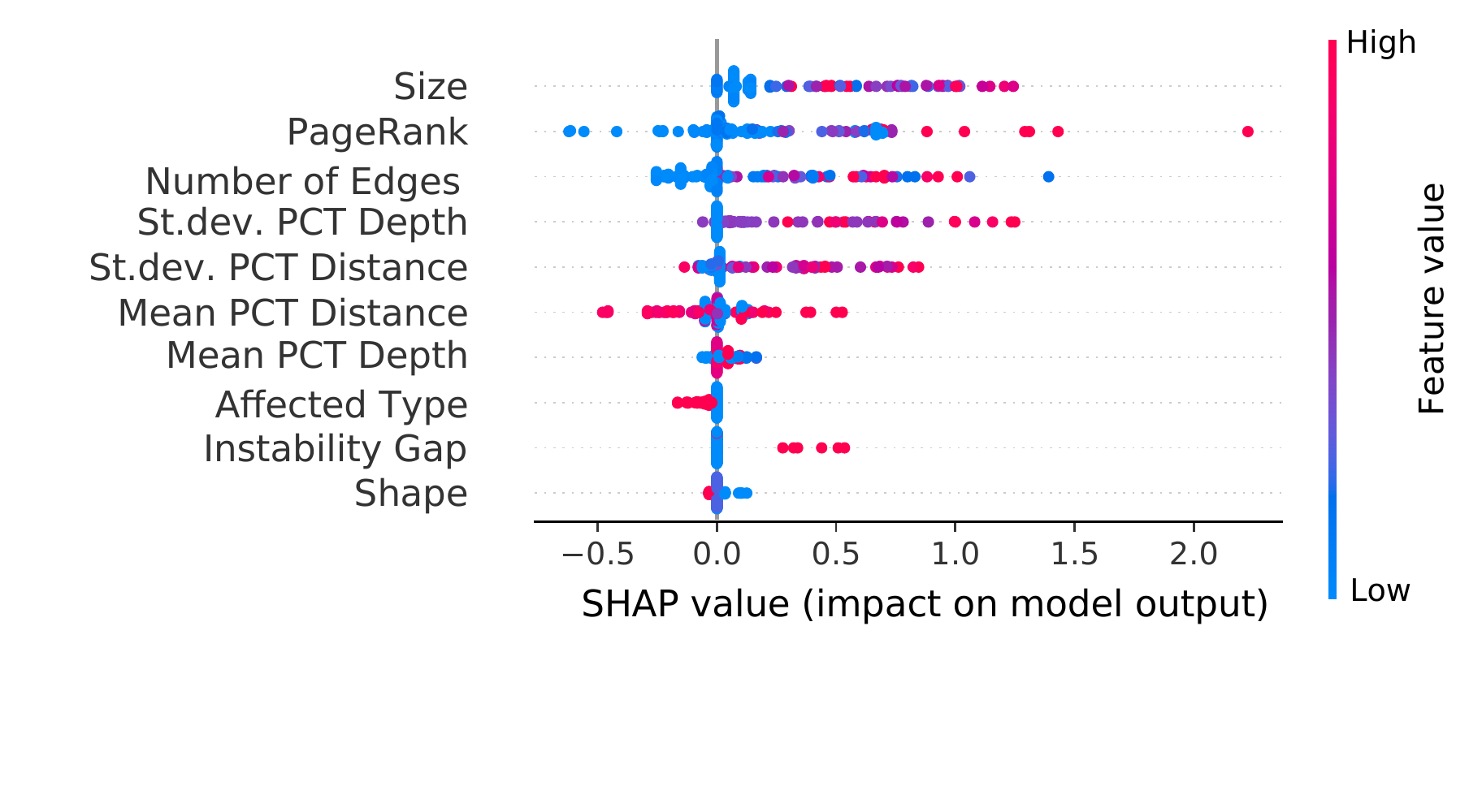}
    \caption{Importance as calculated by the SHAP method \cite{Strumbelj2014}. The higher on the $y$-axis the higher the importance.  Positive values on the $x$-axis mean that the feature contributes to increase the severity of a smell, whereas negative values do the opposite. Colour is mapped to the value assumed by the feature.} 
    \label{fig:feat-explain-shap}
\end{figure}

Figure \ref{fig:feat-explain-shap} depicts the importance of the smell characteristics (or features) according to the SHAP method. Values on the $x$-axis represent the output of the SHAP method: positive values entail that a feature contributed positively (i.e. increased severity) to the output of the model whereas negative values provided a negative contribution (i.e. decreased severity).
We expect the model to match the assumptions in the literature in order to establish that it works as intended.

The \emph{Size} feature (i.e. number of affected elements) contributes the most to the severity of the smell, with higher values of \emph{Size} increasing the severity and small values being neutral.
The \emph{PageRank} of the affected elements comes second, with higher values positively contributing to the severity of a smell.
This means that elements that are more central in the dependency network of the system \textbf{make a smell more severe}.
The \emph{Number of edges} feature has a similar impact as \emph{Size} (they are indeed correlated \cite{Sas2019}), but low values decrease the severity of a smell, instead of being neutral.
The metrics based on the \emph{Package Containment Tree (PCT)} \cite{Laval2012,Al-Mutawa2014} were relatively important too.
High values of \emph{St.dev. PCT Depth}, namely, when a smell affects both elements at the top and bottom of the PCT, positively contributes to increase the predicted severity.
Similar with the \emph{St.dev. PCT Distance}, namely, when a smell affects elements from distant branches of the PCT.
In both cases, it is interesting to note that the mean of both \emph{Distance} and \emph{Depth} are less important than their standard deviation.

After considering these results, we can confirm that the way the model uses the features reflects what is expressed in the literature.
More specifically, the smell gets more severe in cases when its size increases \cite{Lippert2006}, its \emph{centrality} in the dependency network of the system is higher \cite{Roveda2018}; also, under the assumption that distant elements in the PCT are more likely to implement different concerns \cite{Laval2012}, the smell affects elements that are  \emph{unrelated}.

\begin{figure*}
    \centering
    \begin{subfigure}[b]{\linewidth}
        \centering
        \includegraphics[clip, trim=0cm 2cm 0cm 0cm,width=\textwidth]{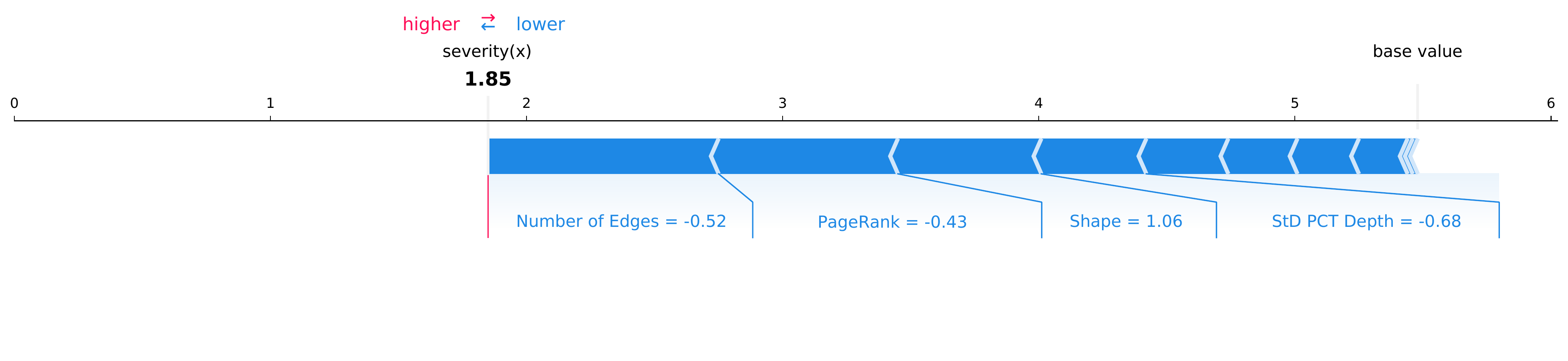}
        \caption{Output explanation of Figure \ref{fig:severity-low}.}
        \label{fig:explain-low}
    \end{subfigure}\\
    \begin{subfigure}[b]{\linewidth}
        \centering
        \includegraphics[clip, trim=0cm 2cm 0cm 0cm, width=\textwidth]{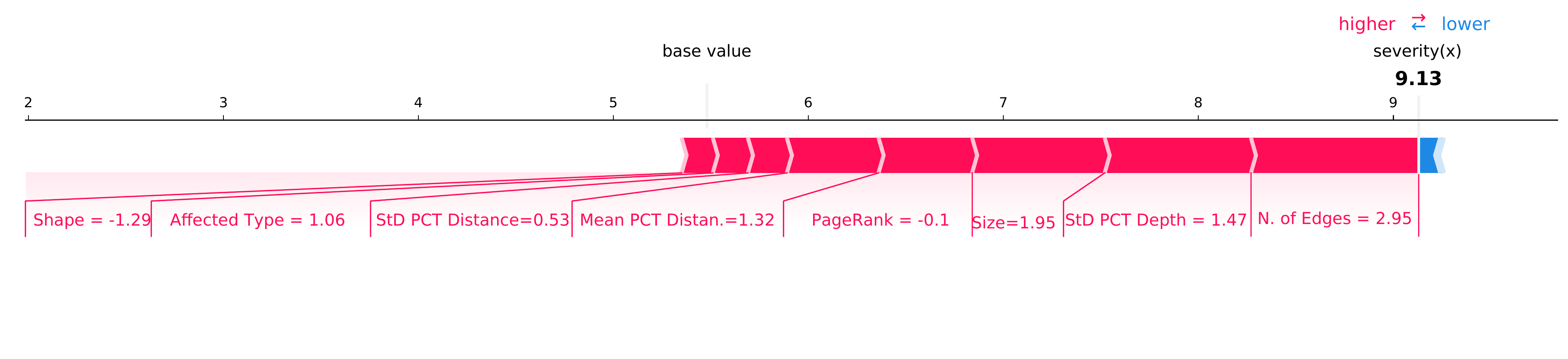}
        \caption{Output explanation of Figure \ref{fig:severity-high}.}
        \label{fig:explain-high}
    \end{subfigure}
    \\
    \begin{subfigure}[b]{\linewidth}
        \centering
        \includegraphics[clip, trim=0cm 2cm 0cm 0cm, width=\textwidth]{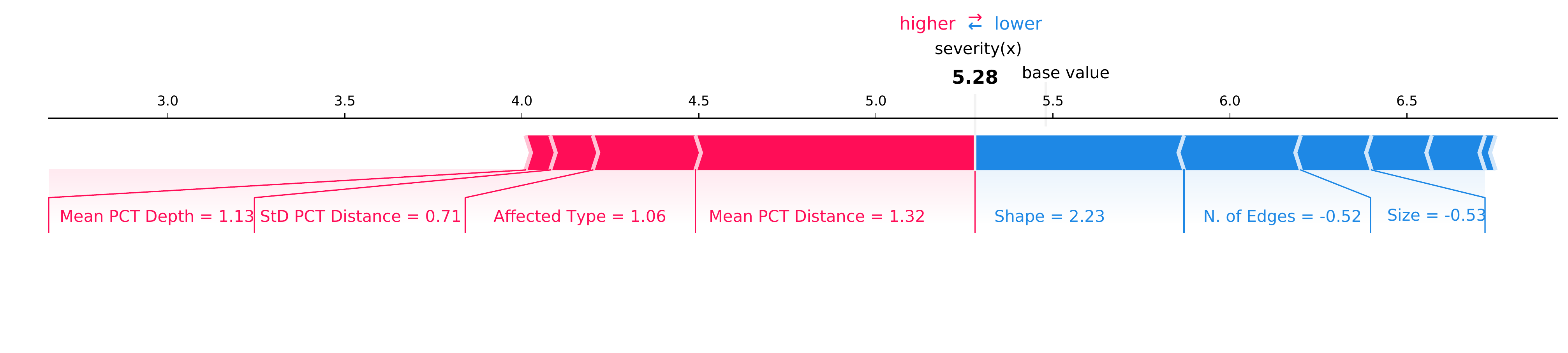}
        \caption{Output explanation of Figure \ref{fig:severity-medium-1}.}
        \label{fig:explain-medium-1}
    \end{subfigure}
    \\
    \begin{subfigure}[b]{\linewidth}
        \centering
        \includegraphics[clip, trim=0cm 2cm 0cm 0cm, width=\textwidth]{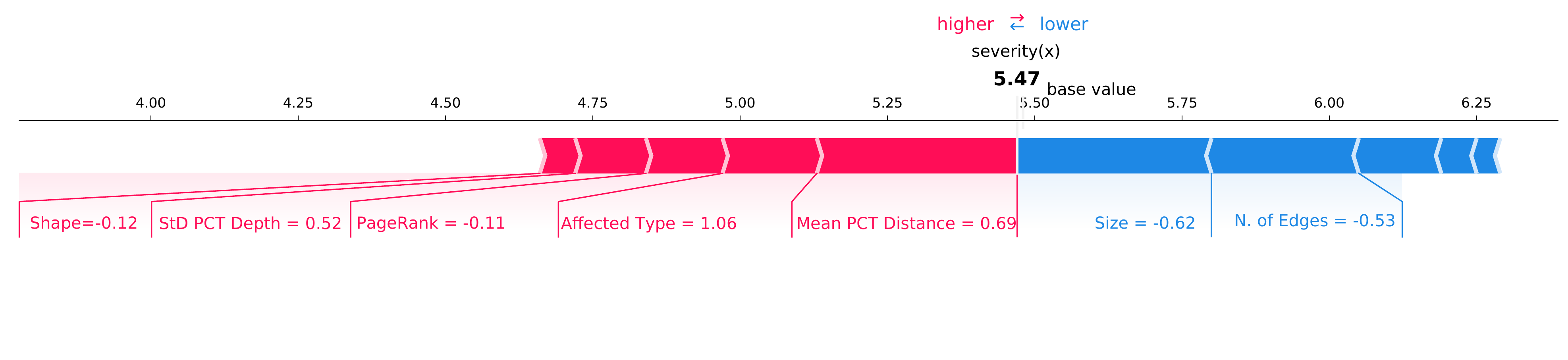}
        \caption{Output explanation of Figure \ref{fig:severity-medium-2}.}
        \label{fig:explain-medium-2}
    \end{subfigure}
       \caption{Prediction explanation of how severity was calculated for smells in Figure \ref{fig:severity-all}. The $x$ axis represents the severity (i.e. output of the model), the blue bars represent a reduction of severity (i.e. negative contribution to prediction). Red bars represent an increase in severity (i.e. positive contribution to prediction). Each segment belongs to a specific feature only. The size of the contribution corresponds to the length of each segment and can be read on the $x$-axis. The value shown next to each feature is the normalised value the feature assumes for the smell instance (i.e. it is \emph{not} the contribution). The number in bold shown above the $x$-axis is the predicted severity for the instance.}
       \label{fig:explain-all}
\end{figure*}

SHAP is also able to explain the output of single instances.
Figure \ref{fig:explain-all} depicts the force plots on how smell characteristics (i.e. features) contributed to the predictions shown in Figure \ref{fig:severity-all}.
For the low-severity smell, Figure \ref{fig:explain-low} shows that  \emph{all features} contributed to reduce the predicted severity of the instance.
The \emph{Number of edges} and \emph{PageRank} features were the two main drivers for the decision.
An almost opposite situation can be observed in Figure \ref{fig:explain-high} for the smell with the highest severity, but in this case the high number of connections (i.e. \emph{Number of edges}) and the fact that smell involves several elements from different parts of the system (i.e. high \emph{Std. dev. PCT Depth}) were the two main drivers behind the prediction of the model. 
Concerning the two smells with similar severity, we can notice in Figures \ref{fig:explain-medium-1} and \ref{fig:explain-medium-2} that the \emph{PCT characteristics} push for a higher severity, but the size-based characteristics push for a lower severity.
These two opposite forces result in a decision that settles towards the middle of the output scale.

In summary, by showing what AS characteristics are used by our model (and how) allows us to better understand what constitutes a severe smell and what does not.
More importantly, it increases the reliability of our study as we do not treat the ML model as a black box. 
Instead, we provide data to explain why it works well and that the identified reasons are in line with what we expected from the literature.

\section{RQ2 results}\label{sec:rq2-results}
In this section we report on the results obtained by analysing the data collected through our interviews.
Note that this section concerns the estimations of the index (as defined by Equation \ref{eq:atdi-smell}), which are calculated using severity (i.e. RQ1 model) but also the extent of the smell.
In the upcoming sections, we first report the opinion of the engineers on the estimations of architectural debt principal (or $ATDI$, the effort to refactor) to answer RQ2.1 and RQ2.2.
Then, we report the general feedback we received from the engineers concerning our approach to answer RQ2.3.
Finally, we conclude by reporting on a few drawbacks and possible improvements of our approach.

\subsection{Perception of the ATDI estimations (RQ2.1 \& RQ2.2)}
\subsubsection{Overview}
Overall, the feedback provided by the participants regarding how well the estimated principal represents the refactoring effort (RQ2.1), was rather positive.
Of the 62 total smell instances that we discussed and their respective ATDI estimations\footnote{Note that for a few participants we did not have time to discuss all 4 instances.}, shown to the participants, 71\% (44/62) of the estimations were described as \emph{\textbf{representative}} of the effort necessary to refactor.
More specifically, responses on industrial instances showed 81\% agreement rate with the estimations provided by the index, whereas for smell instances detected in open source projects the agreement rate with the index was 65\%.
Of the 29\% (18/62) of total instances that were off the mark, only 10 of them were off by more than 100. 
The other 8 instances were off by less than 100, but since 6 of these were small instances, with an $ATDI \le 100$, the relative error was higher, so they were perceived by the participants as a big over-, or under-estimation.
Note that from our descriptive analysis of ATDI (see Figure \ref{fig:atdi-distribution}), we know that only 33\% of instances have an $ATDI \le 100$, so the extent of the imprecision is limited to a small percentage of these 33\% of instances.

Concerning the magnitude and relative size of the estimations (RQ2.2), we established through coding and a typical Likert-scale that 62\% (10/17) of the participants \emph{totally agreed} with the relative size and order of the estimations, while 26\% (4/17) of the participants \emph{disagreed} with the ranking of a single instance only, and the remaining 12\% (2/17) with more than one.
Participants interviewed on industrial projects had a much higher rate of agreement with the ranking and relative size of the estimations than open source participants.
85\% (6/7) of industrial participants completely agreed with the ranking and relative size, whereas only 44\% (4/9) of the open source participants did so.
The rest of the open source participants (5/9) made either one or two corrections to the order.
Note that these were mostly made on instances with an $ATDI < 100$.

The true and false positive rates of the smells presented to our practitioners are the following: 75\% (47/62) of the AS instances were considered true positives by our practitioners, and the remaining 25\% were considered false positives (not seen as a real problem).

The aforementioned numbers provide a quantitative overview of the perception of the participants regarding the validation of ATDI.
In the next sub-section, we will give examples of six different cases, in order to provide a richer, qualitative description of both the smells and the participants perception.  
The examples were chosen to best represent the various aspects of our data set such as: (1) the ratio of agreement/disagreement with estimations; (2) whether the project is open source or industrial; (3) whether it concerned large or small smell instances; and finally, (4) whether the  cases simply presented more insights.

\subsubsection{Example opinions of the participants}\label{sec:examples}
\paragraph{Example 1: \emph{RxJava}} This first example describes how two architectural smells of two different types, GC and HL, are estimated and compared by participant P4.
The GC smell was detected on the package \texttt{io.reactivex.rxjava3.core}, directly containing $52.000$ lines of code spread across 44 classes -- much higher than the average of $10.000$ lines of code circa detected in the other packages of the system.
P4 mentioned that the \texttt{core} package provides access to all the functionality of RxJava through 5 core classes, described as ``god classes''.
For this reason, P4 was exceedingly confident that the estimated value of 1500 for $ATDI$ was \textbf{justified and representative}.

The HL smell was detected on one of the 5 god classes, \texttt{io.reactivex.rxjava3. core.Flowable}, that is part of the GC smell.
This class has an overwhelming number of ingoing and outgoing dependencies, namely 264; in other words, there are 264 other classes that either depend on, or are depended by \texttt{Flowable}.
P4 mentioned that this did not cause any significant technical issue as \texttt{Flowable} does not contain any logic, but it did raise many concerns among the users of RxJava as they lamented the presence of too many methods in this class (as well as in the other 4 god classes).
For these reasons, P4 stated, with great confidence, that the estimated value of 385 for $ATDI$ was \emph{correctly} representing the effort necessary to refactor, and added that it made sense for it to be close to a fifth of the amount estimated for the GC smell as the other four classes shared the same issues and together constitute GC smell itself.

\begin{quote}
    P4: \emph{``[...] this package [the god component] contains, among others, 5 huge classes, which you could consider 5 god classes. So it makes sense to have such a big value for the index.
    There are no real technical issues with it, but users do complain about having too many methods on these god classes.''}
\end{quote}

It is worth mentioning that P4 admitted that every time a new feature was added to the system, these 5 classes were bound to change significantly, as they had to be adapted to the new functionality, as well as updated with the latest Java documentation.

\paragraph{Example 2: \emph{Occupancy of parking facilities}}
This example features a project provided by C2 that monitors occupancy in parking facilities.
The smells discussed for this system were two HL, one affecting a class and the other a package.

The HL at the package level was detected on the \texttt{service} package, the core package of the system containing all the services\footnote{Services are implemented through the Spring Framework.} provided by the system.
The package had a total of 23 ingoing and outgoing dependencies, meaning that it was connected to the majority of the packages in the system.
The estimated $ATDI$ for this smell was $600$.
The \texttt{service} package also contained an HL at class level, namely the \texttt{UserService} class, with 38 ingoing and outgoing dependencies.
This HL smell had an estimated $ATDI$ of 150.
P14 confirmed that estimations for both smells were \textbf{reasonable} as the \texttt{service} package contained the core business functionality of the system (with classes such as \texttt{UserService}), thus making it both very risky (i.e. changes may propagate easily) and very hard to change (i.e. the package is complex because of the business logic).
Moreover, P14 mentioned that \texttt{UserService} was clearly contributing to the \texttt{service} package being an HL as it depended on classes outside \texttt{service} itself, but there were also several other classes contributing to the unbalanced number of dependencies that make \texttt{service} itself a HL.

\begin{quote}
    P14: \emph{``Considering that every single business logic is in there [the \texttt{service} package], yes, I believe that it [the index] is proportionally correct. Especially with respect to the \texttt{UserService} class. The business logic of our services is the most difficult to change, whereas \texttt{UserService} class is relatively easier to change''}.
\end{quote}

\paragraph{Example 3: \emph{Document management system}}
This example features a project provided by C1 affected by several smells.
Among the four smells discussed with P11, the CD is the most interesting to look at due to the counter-intuitive nature of the smell and the estimated ATDI value.

\begin{figure}
    \centering
    \includegraphics[width=.4\linewidth]{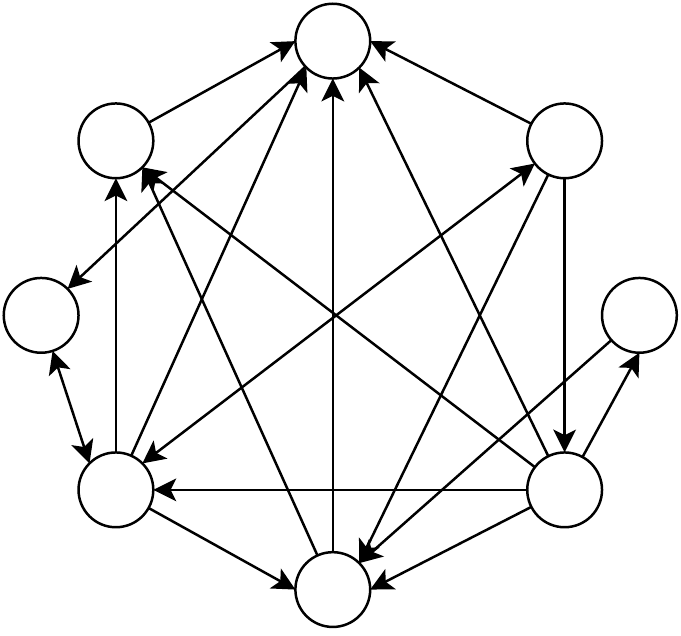}
    \caption{A cycle among 8 classes detected in one of C1's system.}
    \label{fig:c1-docmgmnt-examples}
\end{figure}

The CD is depicted in Figure \ref{fig:c1-docmgmnt-examples} (anonymised to respect the intellectual property of C1) and it affects 8 classes scattered across 6 different packages. 
The classes involved are part of the Model-View-Controller architectural pattern and their purpose is to retrieve data from the database and display it to the user in a view.
Despite these classes being rather intertwined, our model estimated an $ATDI = 65$, a rather low value for eight classes that are so much interconnected. Participant P11 agreed with the estimations, justifying them as follows:

\begin{quote}
    \emph{``The [estimated value] seems pretty good. [...] There are some dependencies that we cannot remove, and I can't see any dependencies that shouldn't be there. So all the dependencies are desired.''}.  
\end{quote}

\paragraph{Example 4: \emph{Jenkins}}
The fourth example features smells detected in the Jenkins project.
Jenkins is a well-known build automation system that has a rather long and convoluted development history.
This resulted in many architectural smells forming in the system over time, two of which are discussed in this example, including the smell with the highest value of ATDI we measured in this study.
The two smells that are of interest are a HL and a GC, which both affect the same package, the \texttt{hudson.model}; this is a huge package that directly contains $43.000$ lines of code distributed across 172 classes with a total of 103 ingoing and outgoing dependencies towards other packages in the system.
The package was described as rather \emph{complex to evolve and change} due to its internal logic and the amount of lines of code.
This example is rather interesting to discuss as the two values of ATDI are quite different from each other despite the two smells affecting the same package.
The estimated index for the HL smell was 8500, whereas for the GC it was 1500.

Nonetheless, P5 \textbf{agreed with great confidence} on both estimations and acknowledged that they were both representative of the effort required to refactor each smell.
P5 provided two reasons on why the refactoring of the HL smell (i.e. reorganise the dependencies to reduce their number) was so difficult.
First, several other parts of the system would have to \emph{change} in order to remove the smell;
and second, \emph{complex refactoring techniques} would be required to do so, mentioning inversion of control and the definition of new APIs as examples.
Both of these do not necessarily hold true -- at least not to the same extent -- for the GC smell: its refactoring would require less invasive operations such as splitting the package into multiple sub-packages.
P5's comment on the matter was that refactoring GC should be easier because its refactoring is more ``self-contained'', that is changing it would impact fewer classes outside the affected package itself.

\begin{quote}
    P5: \emph{``When I think about some of the main things it's [the \texttt{model} package] referring to, most of these things have to do with the build queue logic. And that's the sort of thing that I remember suggesting extracting into a library [...] to untangle the mess within. The idea was that anyone who is much more familiar with the algorithms behind [the job scheduler] and would want to contribute improvements to, would be very unlikely to be able to do so in its current state because of it being a God component. 
    So I definitely agree with the number and agree that it should be a lot lower than the hublike one, particularly because it's more self-contained.''}
\end{quote}

\paragraph{Example 5: \emph{Financial assets management}}
The participants did not always agree with the estimations of ATDI. One such example, is P15, from company C2, who \textbf{disagreed with the estimations} provided for two CD instances discussed during the interview, considering them to be \textbf{overestimated}.
Both cycles affected 3 elements (the first was on packages and the second on classes), which allowed the execution of predicates to filter the trading assets retrieved from a repository according to a certain business logic. 
These two cycles, while unrelated (i.e. in different parts of the system), shared the same logic.
The cycles were estimated at $ATDI = 90$ and $ATDI = 55$ for the package and class cycle respectively; whereas the ideal values for P15 would have been $ATDI = 10$ and $ATDI = 5$.

P15 supports his adjusted estimations by mentioning that the elements in the cycles are not that coupled together and that cycles themselves were introduced intentionally to support a feature.

\begin{quote}
    P15: \emph{``I think that this should be smaller. I've started introducing [these CDs] myself, then everyone else pretty much copy-pasted the design when they created new entities. 
    [...] I've seen the code in these classes, and I know it's really simple to make the change. The predicates packages do not depend on the implementation of the repositories package, so it's just a few lines of code that I have to change. I don't have to make any big architectural change to remove the dependency there.''}
\end{quote}

Given that these cycles were introduced intentionally, it is impossible for our approach to make this distinction.
Arguably, these two smells exist within the system and may cause a problem in the future, and, to some extent, the estimation is justified.
However, we do agree that the estimation should not be that far from the perceived value.

\paragraph{Example 6: \emph{JUnit 5}}
As a final example, we present another case where the participant disagreed with the estimation provided by our model\footnote{Note that 4 examples agreeing with the estimations and 2 disagreeing follows the agreement-disagreement ratio we have in our data (75\%-25\%).}.
This example concerns a GC detected in JUnit on the \texttt{org.junit.jupiter.api} package, which directly contained 54 files, amounting to a total of $9.800$ lines of code\footnote{The detection threshold used for JUnit was $7.800$ lines of code.}.

The ATDI estimated for this smell was 300. P3 was not convinced that this value would be representative of the effort necessary to actually split the package, but it is rather an \textbf{overestimation}.

\begin{quote}
    P3: \emph{``I don't think that splitting that package would be particularly complicated. It's mostly annotations and then assertions and assumptions classes. Those could be relatively easy to split into sensible packages. It's kind of by design and it's the core package, so I'm not in agreement that this is a bad thing in this case. I would agree in general, but maybe this is an exception.''}
\end{quote}

Indeed, the motivations provided by P3 are reasonable, and we accept that the model did overestimate the ATDI in this case. Again, in Section \ref{sec:general-feedback} we discuss how we use this feedback to improve the model.

\subsection{Feedback on the overall approach (RQ2.3)}\label{sec:general-feedback}
In addition to eliciting the perceptions of participants on the ATDI estimations, we also collected some general feedback on the approach. We classified this feedback into three different categories, which are elaborated in the following paragraphs.

\paragraph{Added value}
Most participants (especially the industrial ones) expressed their positive feedback on the added value of adopting architectural smell analysis and a technical debt index.
One of the aspects that was most helpful to most participants was being able to see the smells graphically represented. 
Several participants mentioned that the visual representation of the packages and classes affected by smells can support them in adopting a refactoring strategy to make the components more independent and reusable.
\begin{quote}
    P13: \emph{``Being able to see things visually gives you an insight that we could at least try to structure packages a little differently.''}
\end{quote}

Other participants mentioned the usefulness of the smell detection itself and the estimation of the index specifically when addressing long-standing issues within the project and for prioritisation purposes.

\begin{quote}
    P9: \emph{``[...] breaking down these big, nasty packages is a long-standing issue of the project and a barrier to our ongoing maintenance. Having tools that can automate detection and suggest the index is really nice.''}    
\end{quote}

\begin{quote}
    P2: \emph{``It might be useful in the fact that you can get an estimate of the biggest problem, in this case a god component, and that might be the first to look into.''}
\end{quote}

Finally, the participants acknowledged the value in combining this analysis with continuous integration, as it would help them spot emerging trends and make decisions accordingly on what parts of the system to refactor next.
\begin{quote}
    P5: \emph{``this is the sort of thing that you can calculate on each commit or change and have rules like “you can't increase the technical debt on the project” [...]''}
\end{quote}

\paragraph{Discussion enabler \& learning opportunity}
The participants remarked that, the fact that smells are visualised and assigned an index representing the effort to refactor, eases maintainability-related discussion with the other maintainers of the project.
The ensuing discussion is also objective, as it is backed by the data collected from the current version of the system, rather than by how one specific user, or maintainer, sees the system.

\begin{quote}
    P5: \emph{``[...] seeing the god component there might have been good evidence for when I was trying to suggest [to the other maintainers] the extraction of the queuing and scheduling logic into a library.''}
\end{quote}

In addition to enabling discussion, participants also reported that seeing the smells detected in the code they wrote provided an opportunity for improving themselves because they could understand what mistakes they made.
This would then lead, over time, to personal growth and allow them to write code while also being aware of the architectural implications of their design decisions.
It is noteworthy that some developers were intuitively familiar with the concepts of architectural smells, but they did not a have formal definition to think about them.
\begin{quote}
    P15: \emph{``I think all developers could benefit from something like that. I happen to be a big fan of clean code but I haven't really thought of a clean architecture to be honest.''}
\end{quote}

\paragraph{Limitations of the approach and possible improvements}
The interviewees also allowed us to identify a few limitations and possible improvements to the approach.

Some of the smells were detected on code that did not change in years, and was not expected to change in the future either. 
While these may not be false positives, as the smells were confirmed by engineers to exist, they should be distinguished from other smells, elements of which are constantly changed (and thus TD interest accumulates).
Therefore, we could combine ATDI values with TD interest information that takes into consideration historical change data, and give lower priority to those smells whose affected elements did not change much over the previous years/months.

One challenging issue, is that certain design choices that are detected as architectural smells, do not pose any concern to developers (e.g. all API-related classes are in a single package, like in Example 6).
Therefore, they perceive the ATDI for these instances as an overestimation.
One way to improve that would be to provide more features that allow the model to differentiate between regular classes and abstract classes, interfaces, or annotations. That would allow the developers to tune the model so that their design choices are taken into account in the estimated severity of the smells.

Another limitation is that our approach does not consider the case where cycles formed among classes are caused by interfaces, which are defined much higher in the abstraction hierarchy than the normal classes themselves.
These cycles are much harder to fix because they also require fixing the design of the interfaces. Therefore, the effort required to fix them may be several order of magnitude higher, as it involves changing several other classes.
This aspect results in an underestimation of the ATDI by our approach.
Improving on this aspect would provide much more accurate estimations for smells with small ATDI values.

Finally, some participants expressed their concern with the applicability of the refactoring opportunities suggested by our analysis (smells detection and ATDI) to established libraries and projects sensitive to certain run-time qualities (i.e. reliability and availability).
Architectural smells require large refactorings in order to be changed, and some participants mentioned that it would be hard for them to convince the community to make the necessary changes.

\section{Discussion}\label{sec:discussion}
In this section we discuss the results obtained in our study for each research question as well as their implications for both researchers and practitioners.

\paragraph{General implications} The main implication stemming from our results is that practitioners now have a \emph{validated} approach to measure the ATD principal generated by architectural smells.
This allows them to better track the ATD incurred over time, identify trends in the amount of debt incurred, and react accordingly. 
In particular, this enables them to identify refactoring opportunities and plan them as necessary.
Indeed, AS are well suited for repayment as they are \emph{targeted}, meaning that it is clear for practitioners where the debt is and what steps need to be taken in order to repay it.

Moreover, our approach provides ATD principal estimations for each individual AS smell instance. 
This is instrumental during the prioritisation phase, as practitioners can adopt different prioritisation strategies based on the amount of debt accrued by each instance.
For example, some may decide to refactor the smells with high ATDI to tackle the biggest problems first, whereas others may decide to focus on the small smells only and integrate AS refactoring in their process, resulting in an incremental repayment.
Yet another example was suggested by one of the interview participants: to avoid the introduction of commits that increase the debt over a certain amount, thus resulting in less ATD density over time.

To facilitate the adoption of our approach by industry practitioners, an implementation was integrated into \textsc{Arcan} and is publicly available in the replication package of this study \cite{ReplicationPackage}, or online at \url{https://github.com/Arcan-Tech/arcan-trial}.

\paragraph{RQ1 implications}
For researchers, the main implication stemming from RQ1, is that techniques such as \emph{pairwise comparison}, \emph{ranking systems} (e.g. TrueSkill), and \emph{machine-learned ranking} (or LTR), that are widely adopted in other disciplines such as Information Retrieval (IR), can be flexible enough to be applied to practical problems encountered in Software Engineering (SE).

Indeed, IR complements SE, and more specifically software maintenance, very well.
The core problem faced during software maintenance is the complexity generated by software, namely the difficulty to understand, browse, and change software artefacts because of the high density of information contained in them \cite{Robillard2010}.
IR provides the means to reduce this information overload and only access the information that is needed the most (i.e. the most relevant) based on a given query (e.g. what are the most severe smells in the system?) \cite{Robillard2010}.
Therefore, it comes naturally to think of applying IR techniques to solve SE problems that may be otherwise too complex.
One example of possible application is suggesting the issues (from the issue tracker) that an open source contributor can address based on their previous experience in solving issues and urgency of the issue calculated based on the users' comments on that issue (e.g. new contributors can address low-urgency, low-impact issues).

Alas, applying IR into SE in practice poses some technical challenges that may not be easily overcome in all contexts, or worth the extra effort.
Take for example the problem posed by using pairwise comparison to order a set.
Theoretically, the number of pairwise comparisons that are necessary to obtain a perfectly ordered set grows factorially with the number of elements in the set.
This problem, in fact, arises only to solve another, arguably bigger problem that many IR techniques face: the \emph{need of a data set} to train a machine learning model.
This makes several IR techniques a feasible solution to a ML model, only if a data set already exists, or there is a practical way to create a such data set.
In our case, we were able to create a data set by using pairwise comparison, and subsequently managed to circumvent the problem that pairwise comparison creates by employing several different techniques; but these are clearly limited to our application and may not always be feasible in other contexts.

Nevertheless, we believe that IR applications to SE are quite promising and that there are a lot of potential applications of IR to SE \cite{Happel2008,Robillard2010}.
Compared to the traditional SE approach of designing an algorithm to solve a problem, IR shifts the effort from designing the algorithm to designing the data representing the problem to be solved.
The major drawback is that the lack of means of collecting such data hinders the applicability of such techniques. However, a data-driven solution is more likely to be effective.
In fact, previous studies from the literature have already proven the potential of Recommendation Systems, for example, to suggest design patterns to apply to a code base \cite{Palma2012}, or suggest the libraries to use for a software project and how to use them \cite{DiRocco2021}.
This work has improved on top of that by also demonstrating the potential of ranking systems (e.g. TrueSkill) and machine-learned ranking (LTR).
One concrete idea stemming from our results, is to apply LTR models to create a system that helps developers finding refactorings examples given an AS, or in other words a search engine for refactoring examples.

\paragraph{RQ2 implications}
An interesting remark stemming from the results of RQ2.1 and RQ2.2, is that ATDI \emph{does not need to be precise in order to be considered representative of the effort needed to refactor}.
This is especially true for large estimations (i.e. $ATDI > 500$), because developers seemed to value more the relative value of an estimation (w.r.t. to other estimations) over its absolute value.
Specifically, if in their mind the difference between the largest estimation and the second largest estimation was big enough, then both estimations were deemed representative of the effort. 
However, for smaller instances, some developers (mostly open source ones) did not fully agree with the estimated value of at least one of the instances they were shown.
This shows that the \emph{smaller estimations need to be more precise} in order for them to better resonate with the gut feeling of the engineers and be considered representative.
One way this could be achieved, is by implementing the improvements mentioned in RQ2.3 (Section \ref{sec:general-feedback}).
Nonetheless, given that a large percentage of the maintenance effort is generated by the top few AS \cite{Xiao2016}, we claim that, to a certain extent, our approach provides \emph{\textbf{meaningful}} and \emph{\textbf{representative}} estimations that can be used to both provide objective evidence to make \emph{informed prioritisation decisions} and enable developers to \emph{reliably plan} the allocation of resources during repayment. 

An observation stemming from the results of RQ2.3 is that most projects prioritise other quality attributes over Maintainability/Evolvability.
In our previous work \cite{Sas2020b} on the matter, we found that (industrial) software practitioners prioritise run-time qualities (e.g. Performance, Availability, Evolvability) over design-time qualities (e.g. Maintainability, Evolvability, Compatibility, etc.).
These findings also apply to this study as well, but they also uncover a rather interesting problem that all TD management approaches share.
For projects such as established libraries (e.g. JUnit, RxJava, etc.) and high-availability systems (e.g. Cassandra), \emph{applying refactorings is almost impossible}, as several factors drastically limit the type of changes that the maintainers can make.
For example, moving a class to another package would change its fully qualified name, thus breaking all third party systems depending on it (e.g. think of JUnit).
With a limited amount of options to actually perform ATD repayment, properly managing ATD becomes harder.
It is important to note that this is not specific to our approach, but rather it is a problem faced by all approaches that identify architectural smells and other issues that require large refactorings in order to be removed.

This issue is further aggravated by the fact that many libraries were written before the advent of Java 9 modules and the strong encapsulation they provide. This means that several classes that were only intended to be used internally are instead depended upon by the users of the library, which makes them very hard to change.
Moreover, several senior maintainers of the project are \emph{reluctant to apply any sort of refactoring} for the fear of introducing a bug that would undermine the runtime stability of the project.
Contributors are forced to pay extra TD interest while performing typical maintenance tasks but cannot make the large refactorings required to reduce the amount of interest paid in fear of breaking the backwards compatibility or a key runtime quality of the system. 
This is a \textbf{\emph{deadlock}} situation for maintainers and a \emph{lose-lose} situation for the overall project, as any action undertaken is a risk to the stability of the project.
A possible solution is to embrace API-disruptive changes (e.g. move class to another package) and only include them in major releases of the system.
This strategy comes with the risk of fragmenting the user base and having to maintain two, or more, versions of the same project simultaneously.
 
To conclude, the main implication for practitioners arising from RQ2 is that they can rely on our approach to manage ATD principal, but they might need careful consideration on how to exactly implement it in certain projects that are sensible to change.
Projects that lack proper encapsulation may consider planning for a major release that breaks backwards compatibility, whereas projects that prioritise run-time qualities, may apply small, incremental refactorings that improve the state of the system without compromising its availability or reliability.

\section{Threats to validity}\label{sec:threats-to-validity}
This section describes the threats to validity we identified for this study.
We classified them under \emph{construct validity}, \emph{external validity}, and \emph{reliability}, following the guidelines proposed by Runeson et al. \cite{Runeson2009}.
Internal validity was not considered as we did not examine causal relations \cite{Runeson2009}.

\paragraph*{Construct validity}
This aspect of validity concerns the extent to which this study measures what it is claiming to be measuring \cite{Runeson2009}. 
In other words, whether the data collection and data analysis methodologies truly allow us to answer the research questions we asked.
To ensure that, we developed a case study following a well-known protocol template \cite{Brereton2008}, kept track of how each finding links to the data (chain of evidence) \cite{Runeson2009}, and the study design was reviewed by the two authors iteratively as well as by other researchers within the same research group.

A possible threat to construct validity lies in our selection to use \textsc{Arcan} as the tool to detect architectural smells. Lefever et al. \cite{Lefever2021} have shown that technical debt detection tools report divergent, if not conflicting, results.
This may very well also be the case with \textsc{Arcan} despite not being included in Lefever et al.'s study.
The discrepancy is caused by the fact that different tools adopt different detection rules and provide different implementations of how to detect architectural smells.
Therefore, we cannot claim that the results obtained through \textsc{Arcan} are comparable with results obtained from other tools.
However, it is important to note that this would be the case \emph{even if we used any other tool} \cite{Lefever2021}.
In fact, we also show that this is indeed the case if we use the tool Designite (see replication package \cite{ReplicationPackage}), resulting in the two tools outputting different values of ATDI.
Despite this, the output of ATDI with the two tools is highly correlated (Spearman $\rho = 0.79$), thus meaning that using different tools is unlikely to significantly affect the conclusions drawn (i.e. two projects are likely to be ranked similarly regardless of the tool used.).
Moreover, \textsc{Arcan}'s  the detection rules and algorithms are based on independent, previous work, namely, CD is based on the Acyclic Dependencies Principle \cite{Martin2018,Lippert2006}, HL and UD on the definitions provided by Samarthyam et al. \cite{Samarthyam2016} and Martin \cite{Martin2018}, and GC on Lippert and Roock's principles \cite{Lippert2006}. \textsc{Arcan} was also validated in a number of different studies \cite{Martini2018b,Fontana2020,Fontana2017}.
Therefore, by combining these two aspects (independent detection rules and ATDI correlation with Designite) we consider this threat \emph{mitigated}.

Another threat to construct validity arises from the fact that each participant was asked to discuss (at most) 4 AS instances, and this may not have been enough for them to have a complete impression of the performance of the approach.
This choice was imposed by the limited amount of time we had for each interview (30 minutes), as discussing 4 instances usually required 15 to 20 minutes, and the introduction 10 to 12 minutes.
However, this format gave us the opportunity to discuss the technical details for each instance and better comprehend the point of view, and rationale, of the participants.
This trade-off between quantity and quality allowed us to better motivate our findings and strengthen the chain of evidence.
Therefore, we consider this threat as, at least partially, \emph{mitigated}.

Two more threats to construct validity lie in the selection criteria used to create the training dataset (Table \ref{tab:smells-data-set}) for our machine learning model.
In particular, one could first argue that the minimum number of lines of code is not sufficiently large to guarantee that only non-toy projects were used, and second, that the knowledge of the annotators may have biased the selection excessively.
Concerning the first criterion (i.e. minimum of 10.000 lines of code) we argue that only one project is 10.000 lines of code big, 3 projects have between 23.000 and 31.000 lines of code, and 4 of them have more than 100.000 -- with an average of 115.000.
Concerning the second criterion, we strived to ensure that the projects selected are substantially diverse (web frameworks, parsers, test frameworks, and scientific tools) and, with the exception of two projects (Arcan and AStracker), are relatively well-known open source projects that any other group of annotators could have selected. 
Thus, we argue that the set of projects selected is not sufficiently affected by the specific group of annotators.
Therefore, we consider these two threats as to a large extent \emph{mitigated}.

A final threat to construct validity lies in the method used to sample AS to show to practitioners.
An improper sampling strategy could have caused the sample of smells extracted to mostly focus on smells of a certain type, or on smells of which estimations lie within a specific range (also known as ``cherry-picking'').
This would have inherently biased the results and therefore the outcome of our study.
To avoid such a problem, we adopted \emph{stratified random sampling} to ensure we select the same amount of smells for each smell type (e.g. CD, HL, etc.) while the actual instances sampled for a smell type are picked randomly. 
However, the main problem of this strategy is that it does not reflect the actual ratios of smell types measured in the real world.
Nonetheless, we consider this threat as \emph{mitigated}, as stratified random sampling ensured that the approach is equally validated for all smell types considered while also avoiding ``cherry-picking'' a specific range of values. 

\paragraph*{External validity}
This aspect of validity concerns the extent to which it is possible to generalise the results of the study. In other words, are the results of relevance for cases other than the one analysed \cite{Runeson2009}?

A threat to external validity is the sample of projects used to create the training and test sets for our machine learning model (step `a' in Figure \ref{fig:study-overview}).
More specifically, the pool of projects we sampled smells from, was limited to the projects that our annotators were familiar with. This resulted in the projects (see Table \ref{tab:smells-data-set}) being relatively small (only 2 projects with more than 100.000 lines of code) and the number of application domains covered being relatively limited (3 static analysers, 2 web frameworks, and 3 testing frameworks), on top of all being open source projects.
Ultimately, this could impact the capability of the machine learning model to properly rank AS instances that may be very different than the instances in our training set, both in terms of size and structure.
Nonetheless, we believe that the validation results obtained by RQ2 show that this threat only poses a \emph{limited risk}, as the estimations of the whole approach were not severely impacted despite most of the systems used to validated the results (Tables \ref{tab:open-source-participants} and \ref{tab:industrial-participants}) being very different than the ones contained in the training set (Table \ref{tab:smells-data-set}).

Another threat to external validity is the lack of software architects in our list of participants.
This threat prevents us from claiming that the results of RQ2 can also represent the opinions of software architects, as they may have a completely different opinion on the approach we proposed.
We can, however, claim the generalisation of our results to developers and senior developers with several years of experience.

Finally, the last threat to external validity is the fact that our pool of industrial participants is rather limited. 
We only collaborated with two companies, both are SMEs (Small-medium enterprises) and both are European.
Therefore, it is hard to ensure the full generalisation of the RQ2 results outside these bounds.
However, it is important to notice that almost all of the open source participants were full-time developers in companies from all over the world, and only contributed to the open source project as part of their work, or as a hobby.
Therefore, we believe that this threat is \emph{partially mitigated} and that we can claim the generalisation of our results to full-time open source contributors, industrial practitioners that contribute to open source projects, and (to some extent) industrial practitioners operating in SMEs.

\paragraph*{Reliability}
Reliability is the aspect of validity focusing on the degree to which the data collection and analysis depend on the researchers performing them \cite{Runeson2009}.

While we cannot share the transcription of the interviews for confidentiality reasons, we do, however, provide a replication package \cite{ReplicationPackage} containing the design of this study, the complete list of questions asked to our interview participants, the data set used to train the machine learning model, and the tool \textsc{Arcan} implementing the ATDI calculation.
This should allow researchers to assess the rigour of the study or replicate the results using a different set of projects.

A common threat to reliability when qualitative data is analysed, is the potential bias introduced by the researcher performing the coding.
This threat was mitigated by having a second researcher inspect both the codes and intermediate results during each round of coding. All the feedback received was then integrated and the subsequent coding sessions adopted the updated codes.
The analysis was performed using well-established techniques already used in previous work on the same topic as well as also in different fields (i.e. CCM and Grounded Theory).
Therefore, we consider this threat mitigated.

Another threat to reliability is posed by potential improper application of machine learning techniques during the model engineering phase (i.e. data set creation and model evaluation). 
To avoid common pitfalls when implementing machine learning into our approach, 
the whole process was supervised by a third researcher specialised in this field.
Thus, we consider this threat, at least partially, mitigated.

\section{Related work}\label{sec:related-work}
In this section we report on previous research on topics related to this study, i.e. to the estimation of the principal of architectural technical debt. Specifically, we review related work on the following two categories: approaches estimating \emph{architectural} TD principal (Section \ref{sec:related-work-atd}), and approaches estimating any other type of TD principal (Section \ref{sec:related-work-td}).
Our approach is directly comparable only to the first category, i.e. to similar work on ATD principal estimation; this comparison is presented at the end of Section \ref{sec:related-work-atd}.

\subsection{Approaches estimating architectural debt principal}\label{sec:related-work-atd}
Xiao et al. \cite{Xiao2016} have proposed a formal model to quantify architectural debt principal using a History Coupling Probability (HCP) matrix.
The HCP matrix is calculated using conditional probability of changing files and is then used to identify candidates of debt items.
Candidates were then modelled using different regression models (linear, logarithmic, etc.) to find which model fits best the interest of the debt items in the system.
Next, debt items were ranked based on the effort required to fix them.
Xiao et al. evaluated their approach on 7 open source projects, showing that a significant proportion (51\% to 81\%) of the overall maintenance effort was consumed by paying interest on architectural debt items.
Their approach is implemented into their tool called DV8.

Roveda et al. \cite{Roveda2018} developed an architectural technical debt index based on the (dependency-based) architectural smells detected in the system.
The index is based on the number of smells detected in the system, their severity, the history of the smell in the system, and a few dependency metrics defined by Martin \cite{Martin2018}.
The calculation of severity takes into consideration the PageRank of the architectural smell calculated on the dependency graph of the system.
Next, Roveda et al. proceeded to analyse the Qualitas Corpus data set \cite{Terra2013,Tempero2010} and compare the results with Sonarqube's technical debt index.
The comparison showed that there is no correlation in the historical trends of the two indexes, leading the authors to conclude that the two indexes are independent.

Wu et al. \cite{Wu2018} created and validated an architectural debt index within a big multinational software company.
The index is called Standard Architecture Index (SAI) and is composed of a number of measures reflecting recurring architecture problems reported by the company's engineers and architects.
More specifically, measures are based on coupling, cohesion, rate of cyclic dependencies, instability, modularity violation rate, and many others.
The index went through two major iterations within the company and the authors also compared it to actual productivity measurements.
The improvements measured by SAI correlated with improvements in productivity for the two products the index was tested on.

Martini et al. \cite{Martini2018} proposed a semi-automated approach to identify and estimate the architectural debt principal of a project owned by a large telecom company that is written in C++.
The approach features a measurement system based on the ISO/IEC 15939:2007 standard to estimate the urgency for refactoring for each component in the system based on two key concepts: current complexity of the system and effort spent maintaining the system.
To calculate these, several metrics and algorithms were taken into consideration by the authors, such as the number of files, the number of lines of code, the number of changes in all files, McCabe's and Halstead's complexity metrics.
For the calculation of the effort, however, engineers need to be involved, thus making the process semi-automated.in this regarded with their model.
The results showed that the engineers agreed that the output of the model was useful to identify any architectural debt that needs refactoring and that the effort estimation estimates correctly the business value of doing such a refactoring.

Verdecchia et al. \cite{Verdecchia2020} proposed a generalised approach to calculate  the technical debt principal index of a system leveraging statistical analysis.
Unlike the aforementioned approaches, Verdecchia et al. aimed at designing a process that is language-independent, tool-independent, supports tool composability with multiple levels of granularity of their analysis (e.g. class, package, module, etc.).
To achieve such a goal, they formalised the problem mathematically and considered the output of any tool as a set of architectural rules that are applied to every artefact in the system. 
Next, they incorporated into the mathematical model granularity levels and clusters of architectural rules (called architectural dimensions by the authors).
While this approach does have several advantages (as mentioned above), it also comes with a number of drawbacks: for example, it is dependant on a benchmark of software projects (which has to be continuously updated) to calculate some of the statistics used during the calculation of the index.
The authors evaluated the validity of the approach in a subsequent study using questionnaires \cite{Verdecchia2022}.

Table \ref{tab:rw-comparison} summarises the differences between related work on ATD principal estimation and this work. 
Our work improves over related work on two main points.
First, we are the first to propose a learning-to-rank machine learning-based approach to tackle this problem rather than relying on manually-set thresholds \cite{Wu2018,Martini2018} or arbitrary proxies of severity \cite{Roveda2018}.
Second, we validated our approach by interviewing software developers from both industrial and open source projects; most other approaches were only \emph{evaluated}, in the sense of performing measurements and comparisons, while only two were \emph{validated}, in the sense of confirming they actually provide benefits to developers, but in a limited way.
Specifically, evaluation was mostly performed on open source systems \cite{Xiao2016,Roveda2018,Verdecchia2020}, whereas the only two studies that performed a validation were in both cases within a single company\footnote{Note that several of the open source engineers that we interviewed were also working in industry.} \cite{Martini2018,Wu2018}, thus critically reducing the generalisation of the results obtained. 
Additionally, our approach is the only one that provides a publicly-available tool that implements the approach \cite{ReplicationPackage}.

The use of the smell characteristics of an architectural (or code) smell to predict its severity is certainly not a new concept in software engineering \cite{Laval2012,Tsantalis2011,ArcelliFontana2015,Vidal2016,Roveda2018}.
Some work has focused on the use of metrics by selecting arbitrary, hand-picked thresholds, or weights, and combining such metrics into a single value representing the severity of a smell \cite{Laval2012,Vidal2016}.
Others experimented with using benchmarks of open source systems to automatically define thresholds \cite{ArcelliFontana2015} with the hope of reducing the bias introduced by hand-picked metrics. 
Alas, both of these strategies are inherently flawed. In the former case, hand-picked thresholds, even if based on heuristics and expertise, are severely limited to specific cases dictated by the assumptions (e.g. how much a design principle influences a smell) used to set them in the first place.
Instead, our ML model deduces these from the training set as part of the training process.
In the latter case, benchmark-based approaches assume that the systems included in the benchmark cover the whole spectrum of good, medium, and low quality systems and that the metrics computed on them are distributed equally for `good' and `bad' values of the metric itself (e.g. the Complexity metric only has values for $>.50$ in the benchmark, so $.50$ is considered the lowest value of Complexity when in reality it might be a medium value).


To conclude, the approach presented in this paper does not suffer from some of the well-known shortcomings that other studies do \cite{Khomyakov2020}.
In particular, we developed a fully-automated approach that does not rely on hand-picked thresholds, or benchmarks, but instead uses machine learning to overcome these shortcomings; then, we validated the approach by involving practitioners from multiple companies and from both open source and industry.
In addition, an implementation of our approach is also freely available in the replication package of this paper \cite{ReplicationPackage}.

\begin{sidewaystable}[]
    \footnotesize
    \centering
    \caption{Comparison of related work with the approach proposed in this paper (*Absolute: measurement is dependant only on the input; Relative: measurement depends on both a benchmark (or ML model) and on the input).}
    \label{tab:rw-comparison}
    \begin{tabular}{r|m{2.5cm}m{2.5cm}m{2.5cm}m{2.5cm}m{2.5cm}m{2.5cm}}
    \hline
    \textbf{Related Work} & This work & \cite{Xiao2016} & \cite{Roveda2018} & \cite{Martini2018} & \cite{Wu2018} & \cite{Verdecchia2020} \\
    \textbf{Input} & Source code and architectural smells & Historical change data, Source code, bug trackers, and architectural smells & Java Bytecode, change data, and architectural smells & UML model. Source code and custom thresholds & Historical change data, source code, and questionnaires & Results of 3rd party tools \\
    \textbf{Technique} & ML to estimate severity via smell characteristics and lines of code creating smell to estimate refactoring complexity & Regression models of number of changes, bugs reported, and lines of code committed per file & PageRank as proxy for severity, number of dependencies of the smell, and historical changes & Aggregation of metrics with heuristically-set thresholds & Aggregation of metrics using arbitrary thresholds based on opinion of engineers & Algorithmic approach based on generalised rules for 3rd party tools' violations and a benchmark of systems \\
    \textbf{Measurement*} & Relative & Absolute & Relative & Absolute & Relative & Relative \\
    \textbf{Output} & ATD principal as an index & ATD interest paid so far & ATD principal as an index & ATD principal as an index & Urgency to refactor a component and modularisation index & ATD principal as an index \\
    \textbf{Automation} & Full & Full & Full & Partial & Full & Full but depends on 3rd party tool \\
    \textbf{Availability} & Yes & No & Yes & Yes & No & Yes \\
    \textbf{Validation or Evaluation} & Validated through interviews with both open source and professional engineers & Evaluated on open source systems & Evaluated on open source systems & Validation within a company & Validation within a company & Evaluated \\ \hline
    \end{tabular}
\end{sidewaystable}

\subsection{Approaches estimating other types of technical debt principal}\label{sec:related-work-td}
Letouzey et al. \cite{Letouzey2010} designed the well-known SQALE analysis model that hierarchically aggregates from rough low-level measurements into a high-level index that is meant to represent the status of the whole system.
More specifically, the authors describe how to aggregate the data coming from two different types of hierarchies: an artefact hierarchy (e.g. from lines of code to methods, from methods to classes, etc.) and a quality hierarchy (e.g. Maintainability is broken down into Readability, Changeability, etc.).
Additionally, they also analysed how different data scales (e.g. nominal, ratio, interval, etc.) should be synthesised and aggregated.
SQALE was later adopted and evolved by several tools, including SonarQube\footnote{Visit \url{https://www.sonarqube.org/}.} and SQuORE\footnote{Visit \url{https://www.squoring.com/}.}.

Nugroho et al. \cite{Nugroho2011}, presented a technical debt principal and interest estimation approach.
Their approach uses a straightforward (linear) mathematical model to map ISO/IEC 9126 quality attributes to a series of source code properties.
Next, these properties were mapped to a rating system with 5 different levels (i.e. star-based system) to represent the current quality level of the system.
The approach thus allows to estimate the current principal by calculating the amount of effort required to rework the system to get a higher quality rating (e.g. from 3 stars to 5 stars).
Similarly, the interest is calculated by using the maintenance effort at a given quality level and then subtracting the maintenance effort at the desired quality level.
After describing the approach, Nugroho et al. proceeded to evaluate it on an 18-year-old system that is written in multiple programming languages and has over 760.000 lines of code.
The case study was designed with the goal of illustrating that the proposed approach could be applied to answer practical questions related to software quality improvement over 10 years of development of the said project.
Using the proposed approach, Nugroho et al. showed that 75\% of the system needs to be reworked in order to meet the ideal quality level (5 stars).

Marinescu \cite{Marinescu2012} proposed a technical debt index based on design flaws, which include most of the code smells identified by Fowler and Beck \cite{Fowler2002}.
Marinescu assigned to each design flaw (1) a degree to which it influences coupling, cohesion, complexity, and encapsulation; a (2) granularity (e.g. class, method, etc.), and (3) a metric that influences their severity (e.g. length of duplicated code for the duplicated code design flaw).
The overall score is then computed by aggregating the impact score of each design flaw detected in the system and normalising by total lines of code in the system.
The approach was evaluated on two well-known Java systems, allowing to derive insights on their evolution and on the parts affected by design flaws.
In particular, Marinescu established that several types of flows degraded over time, thus demonstrating the practicality of the approach in real-world scenarios.

In their 2012 paper, Curtis et al. \cite{Curtis2012} report the approach adopted by CAST's Application Intelligence Platform to estimate technical debt principal.
The approach hierarchically divides Maintainability in multiple quality attributes according to ISO/IEC 9126 and ISO/IEC 25010.
At the bottom of the hierarchy, there are up to 506 quality rules, and each rule may be evaluated for more than one quality attribute.
The cost to fix all the violations (i.e. the principal) is then calculated by assigning to each rule a high, medium, or low severity, and then multiplying it by the average number of hours needed to fix each type (e.g. low severity requires fewer hours).
Curtis et al. also described their experience analysing 700 applications and measuring technical debt three times, each time at a different point in the application's history.
The results suggest that the analysis and measurement of technical debt principal using the proposed approach can be used in conjunction with structural quality priorities to guide management decisions regarding future resource allocation.

Mayr et al. \cite{Mayr2014} proposed a classification scheme that enables systematic categorisation of technical debt-estimating approaches.
Moreover, they also propose their own approach based on a benchmarking-oriented calculation of the technical debt principal.
Similarly to other approaches, the approach of Mayr et al. uses a set of rules and abstraction dimensions.
Contrary to other approaches, however, they also rely on a benchmark of systems to create a baseline quality level.
The baseline is simply the distribution of different rules in the benchmark systems.
The remediation cost (i.e. the principal) is then calculated as a linear function of the number of violations to be fixed, the effort required for each violation, and the cost rate.
This approach was evaluated on two open source projects; the results show that the approach was able to provide stakeholders with the expected remediation costs depending on the actual quality of the project and target level.

While all the aforementioned approaches estimate TD principal, they mostly focus on code debt, therefore, they are not directly comparable to our work.

\section{Conclusion and future work}\label{sec:conclusion-fw}
The goal of this work was to estimate the architectural debt principal using architectural smells as a input.
For this purpose, we designed an approach that relies on machine learning and static analysis of the source of the smell to estimate the effort necessary to refactor a smell.
Next, we created a data set to rank architectural smells by their severity using well-known techniques typically used in information retrieval (e.g. TrueSkill).
Then, we trained a type of ML model typically used in information retrieval (called learning-to-rank) and obtained excellent results, thus demonstrating that it is possible to apply these techniques in software engineering.
Finally, we validated the output of the whole approach (not only of the ML model), through a case study where we interviewed 16 engineers from both open source and industry.
The results showed that most of the estimations ($\ge 70$\%) provided by our approach are representative of the effort necessary to refactor a smell.
The results also suggested that for large estimations our approach was very precise; however, for some cases, smaller estimations were not as precise.
We also identified several points of improvement for our approach, such as taking into consideration the class hierarchy, as it could have a big influence on the estimations provided by the approach (especially the smaller ones).

Overall, the results confirm that the estimations provided by our approach are, for the most part, in line with the effort estimations expected by industry practitioners.
This means that our approach is a viable option that could allow practitioners to track ATD principal of a system, plan remediation strategies, and prioritise individual AS instances.

Concerning future work, we identified several opportunities. The first one is to add support for C/C++ systems by expanding the data set of the ML model with C/C++ systems, as \textsc{Arcan} already supports the detection of AS for these systems.
Another future work opportunity is to improve the estimations as highlighted in Section \ref{sec:general-feedback}, namely extend the ML model with more features and consider special cases such as cycles in the class hierarchy layers.
Finally, we plan to complement the estimation of the principal with the estimation of the interest -- i.e. the cost of maintaining the current solution -- by exploiting change metrics.
This was also requested by some of our interviewees who understood that in order to take a refactoring decision, they also require to know the cost of keeping the system as is.

\section*{Acknowledgements}
This work was supported by the ITEA3 research project under grant agreement No. 17038 VISDOM.

A special thanks to Cezar Sas for helping us better implement and use the machine learning techniques adopted in this paper.
\bibliography{bibliography}

\begin{thebibliography}{10}

\bibitem{Avgeriou2016}
P.~Avgeriou, P.~Kruchten, I.~Ozkaya, and C.~Seaman, ``{Managing Technical Debt
  in Software Engineering (Dagstuhl Seminar 16162)},'' {\em Dagstuhl Reports},
  vol.~6, no.~4, pp.~110--138, 2016.

\bibitem{Cunningham1992}
W.~Cunningham, ``{The WyCash Portfolio Management System},'' {\em SIGPLAN OOPS
  Mess.}, vol.~4, pp.~29--30, dec 1992.

\bibitem{Ernst2015}
N.~A. Ernst, S.~Bellomo, I.~Ozkaya, R.~L. Nord, and I.~Gorton, ``{Measure it?
  Manage it? Ignore it? software practitioners and technical debt},'' in {\em
  Proceedings of the 2015 10th Joint Meeting on Foundations of Software
  Engineering - ESEC/FSE 2015}, ESEC/FSE 2015, (New York, New York, USA),
  pp.~50--60, ACM Press, 2015.

\bibitem{Khomyakov2020}
I.~Khomyakov, Z.~Makhmutov, R.~Mirgalimova, and A.~Sillitti, ``{An Analysis of
  Automated Technical Debt Measurement},'' in {\em Lecture Notes in Business
  Information Processing}, vol.~378 LNBIP, pp.~250--273, Springer, Cham, may
  2020.

\bibitem{Avgeriou2021}
P.~C. Avgeriou, D.~Taibi, A.~Ampatzoglou, F.~{Arcelli Fontana}, T.~Besker,
  A.~Chatzigeorgiou, V.~Lenarduzzi, A.~Martini, A.~Moschou, I.~Pigazzini,
  N.~Saarimaki, D.~D. Sas, S.~S. {De Toledo}, and A.~A. Tsintzira, ``{An
  Overview and Comparison of Technical Debt Measurement Tools},'' {\em IEEE
  Software}, vol.~38, pp.~61--71, may 2021.

\bibitem{Lippert2006}
S.~R. {Martin Lippert}, {\em {Refactoring in Large Software Projects:
  Performing Complex Restructurings Successfully}}.
\newblock 2006.

\bibitem{Garcia2009}
J.~Garcia, P.~Daniel, G.~Edwards, and N.~Medvidovic, ``{Identifying
  Architectural Bad Smells},'' in {\em Proceedings of the European Conference
  on Software Maintenance and Reengineering, CSMR}, pp.~255--258, 2009.

\bibitem{Verdecchia2018}
R.~Verdecchia, I.~Malavolta, and P.~Lago, ``{Architectural Technical Debt
  Identification: the Research Landscape},'' in {\em 2018 ACM/IEEE
  International Conference on Technical Debt}, 2018.

\bibitem{Fontana2020}
F.~{Arcelli Fontana}, F.~Locatelli, I.~Pigazzini, and P.~Mereghetti, ``{An
  Architectural Smell Evaluation in an Industrial Context},'' no.~c,
  pp.~68--74, 2020.

\bibitem{Sas2021b}
D.~Sas, I.~Pigazzini, P.~Avgeriou, and F.~A. Fontana, ``{The Perception of
  Architectural Smells in Industrial Practice},'' {\em IEEE Software}, vol.~38,
  pp.~35--41, nov 2021.

\bibitem{Xiao2016}
L.~Xiao, Y.~Cai, R.~Kazman, R.~Mo, and Q.~Feng, ``{Identifying and quantifying
  architectural debt},'' in {\em Proceedings - International Conference on
  Software Engineering}, vol.~14-22-May-, pp.~488--498, IEEE Computer Society,
  2016.

\bibitem{Roveda2018}
R.~Roveda, F.~A. Fontana, I.~Pigazzini, and M.~Zanoni, ``{Towards an
  architectural debt index},'' in {\em Proceedings - 44th Euromicro Conference
  on Software Engineering and Advanced Applications, SEAA 2018}, pp.~408--416,
  IEEE, aug 2018.

\bibitem{Jarvelin2002}
K.~J{\"a}rvelin and J.~Kek{\"a}l{\"a}inen, ``Cumulated gain-based evaluation of
  ir techniques,'' {\em ACM Transactions on Information Systems (TOIS)},
  vol.~20, no.~4, pp.~422--446, 2002.

\bibitem{Strumbelj2014}
E.~{\v{S}}trumbelj and I.~Kononenko, ``Explaining prediction models and
  individual predictions with feature contributions,'' {\em Knowledge and
  information systems}, vol.~41, no.~3, pp.~647--665, 2014.

\bibitem{Fontana2016}
F.~A. Fontana, R.~Roveda, and M.~Zanoni, ``{Technical Debt Indexes Provided by
  Tools: A Preliminary Discussion},'' in {\em Proceedings - 2016 IEEE 8th
  International Workshop on Managing Technical Debt, MTD 2016}, pp.~28--31,
  Institute of Electrical and Electronics Engineers Inc., dec 2016.

\bibitem{Martin2018}
R.~C. Martin, J.~Grenning, and S.~Brown, {\em Clean architecture: a craftsman's
  guide to software structure and design}.
\newblock Prentice Hall, 2018.

\bibitem{Fowler2002}
M.~Fowler and K.~Beck, {\em {Refactoring: Improving the Design of Existing
  Code}}.
\newblock Addison-Wesley, 1~ed., 2002.

\bibitem{Fontana2019}
F.~{Arcelli Fontana}, V.~Lenarduzzi, R.~Roveda, and D.~Taibi, ``{Are
  architectural smells independent from code smells? An empirical study},''
  {\em Journal of Systems and Software}, vol.~154, pp.~139--156, 2019.

\bibitem{Sas2019}
D.~Sas, P.~Avgeriou, and F.~{Arcelli Fontana}, ``{Investigating instability
  architectural smells evolution: an exploratory case study},'' in {\em 35th
  International Conference on Software Maintenance and Evolution},
  pp.~557--567, IEEE, sep 2019.

\bibitem{Fontana2017}
F.~{Arcelli Fontana}, I.~Pigazzini, R.~Roveda, D.~Tamburri, M.~Zanoni, and
  E.~D. Nitto, ``{Arcan: A tool for architectural smells detection},'' {\em
  Proceedings - 2017 IEEE International Conference on Software Architecture
  Workshops, ICSAW 2017: Side Track Proceedings}, pp.~282--285, 2017.

\bibitem{Parnas1979}
D.~L. Parnas, ``Designing software for ease of extension and contraction,''
  {\em IEEE transactions on software engineering}, no.~2, pp.~128--138, 1979.

\bibitem{Stevens1974}
W.~P. Stevens, G.~J. Myers, and L.~L. Constantine, ``Structured design,'' {\em
  IBM Systems Journal}, vol.~13, no.~2, pp.~115--139, 1974.

\bibitem{Fontana2015}
F.~Arcelli~Fontana, V.~Ferme, M.~Zanoni, and A.~Yamashita, ``{Automatic metric
  thresholds derivation for code smell detection},'' {\em International
  Workshop on Emerging Trends in Software Metrics, WETSoM}, vol.~2015-Augus,
  pp.~44--53, 2015.

\bibitem{Pawlak2015}
R.~Pawlak, M.~Monperrus, N.~Petitprez, C.~Noguera, and L.~Seinturier, ``{Spoon:
  A Library for Implementing Analyses and Transformations of Java Source
  Code},'' {\em {Software: Practice and Experience}}, vol.~46, pp.~1155--1179,
  2015.

\bibitem{ReplicationPackage}
``Replication package zip file.''
  \url{https://dx.doi.org/10.6084/m9.figshare.19823323}.
\newblock Accessed: 2022-05-23.

\bibitem{Laval2012}
J.~Laval, J.-R. Falleri, P.~Vismara, and S.~Ducasse, ``{Efficient Retrieval and
  Ranking of Undesired Package Cycles in Large Software Systems.},'' {\em The
  Journal of Object Technology}, vol.~11, p.~4:1, apr 2012.

\bibitem{Al-Mutawa2014}
H.~A. Al-Mutawa, J.~Dietrich, S.~Marsland, and C.~McCartin, ``{On the shape of
  circular dependencies in java programs},'' in {\em Proceedings of the
  Australian Software Engineering Conference, ASWEC}, pp.~48--57, IEEE, apr
  2014.

\bibitem{Murillo2021}
M.~I. Murillo, A.~Pacheco, G.~L{\'{o}}pez, G.~Mar{\'{i}}n, and J.~Guzm{\'{a}}n,
  ``{Common Causes and Effects of Technical Debt in Costa Rica: InsighTD Survey
  Replication},'' {\em Proceedings - 2021 47th Latin American Computing
  Conference, CLEI 2021}, 2021.

\bibitem{Rios2020}
N.~Rios, L.~Mendes, C.~Cerdeiral, A.~P.~F. Magalh{\~{a}}es, B.~Perez,
  D.~Correal, H.~Astudillo, C.~Seaman, C.~Izurieta, G.~Santos, and R.~{Oliveira
  Sp{\'{i}}nola}, ``{Hearing the Voice of Software Practitioners on Causes,
  Effects, and Practices to Deal with Documentation Debt},'' {\em Lecture Notes
  in Computer Science (including subseries Lecture Notes in Artificial
  Intelligence and Lecture Notes in Bioinformatics)}, vol.~12045 LNCS,
  pp.~55--70, mar 2020.

\bibitem{Rios2018}
N.~Rios, R.~{Oliveira Sp{\'{i}}nola}, M.~Mendon{\c{c}}a, and C.~Seaman, ``{The
  Most Common Causes and Effects of Technical Debt: First Results from a Global
  Family of Industrial Surveys},'' {\em Proceedings of the 12th ACM/IEEE
  International Symposium on Empirical Software Engineering and Measurement},
  vol.~18, 2018.

\bibitem{Ampatzoglou2018}
A.~Ampatzoglou, A.~Michailidis, C.~Sarikyriakidis, A.~Ampatzoglou,
  A.~Chatzigeorgiou, and P.~Avgeriou, ``{A framework for managing interest in
  technical debt: An industrial validation},'' in {\em Proceedings -
  International Conference on Software Engineering}, 2018.

\bibitem{Letouzey2010}
J.~L. Letouzey and T.~Coq, ``{The SQALE analysis model an analysis model
  compliant with the representation condition for assessing the quality of
  software source code},'' in {\em Proceedings - 2nd International Conference
  on Advances in System Testing and Validation Lifecycle, VALID 2010},
  pp.~43--48, 2010.

\bibitem{Curtis2012}
B.~Curtis, J.~Sappidi, and A.~Szynkarski, ``{Estimating the principal of an
  application's technical debt},'' {\em IEEE Software}, vol.~29, pp.~34--42,
  nov 2012.

\bibitem{Marinescu2012}
R.~Marinescu, ``{Assessing technical debt by identifying design flaws in
  software systems},'' {\em IBM Journal of Research and Development}, vol.~56,
  pp.~9:1--9:13, sep 2012.

\bibitem{Chatzigeorgiou2015}
A.~Chatzigeorgiou, A.~Ampatzoglou, A.~Ampatzoglou, and T.~Amanatidis,
  ``{Estimating the breaking point for technical debt},'' in {\em 2015 IEEE 7th
  International Workshop on Managing Technical Debt, MTD 2015 - Proceedings},
  pp.~53--56, Institute of Electrical and Electronics Engineers Inc., nov 2015.

\bibitem{Kamei2016}
Y.~Kamei, E.~Maldonado, E.~Shihab, and N.~Ubayashi, ``{Using analytics to
  quantify the interest of self-admitted technical debt},'' in {\em CEUR
  Workshop Proceedings}, vol.~1771, pp.~68--71, 2016.

\bibitem{Nugroho2011}
A.~Nugroho, J.~Visser, and T.~Kuipers, ``{An empirical model of technical debt
  and interest},'' in {\em Proceedings - International Conference on Software
  Engineering}, (New York, New York, USA), pp.~1--8, ACM Press, 2011.

\bibitem{Morasca2001}
S.~Morasca and G.~Russo, ``{An empirical study of software productivity},''
  {\em Proceedings - IEEE Computer Society's International Computer Software
  and Applications Conference}, pp.~317--322, 2001.

\bibitem{Kitchenham2004}
B.~Kitchenham and E.~Mendes, ``{Software productivity measurement using
  multiple size measures},'' {\em IEEE Transactions on Software Engineering},
  vol.~30, pp.~1023--1035, dec 2004.

\bibitem{ArcelliFontana2017}
F.~{Arcelli Fontana} and M.~Zanoni, ``{Code smell severity classification using
  machine learning techniques},'' {\em Knowledge-Based Systems}, vol.~128,
  pp.~43--58, jul 2017.

\bibitem{ArcelliFontana2015}
F.~{Arcelli Fontana}, V.~Ferme, M.~Zanoni, and R.~Roveda, ``{Towards a
  prioritization of code debt: A code smell Intensity Index},'' in {\em 2015
  IEEE 7th International Workshop on Managing Technical Debt, MTD 2015 -
  Proceedings}, pp.~16--24, Institute of Electrical and Electronics Engineers
  Inc., nov 2015.

\bibitem{Vidal2016}
S.~A. Vidal, C.~Marcos, and J.~A. D{\'{i}}az-Pace, ``{An approach to prioritize
  code smells for refactoring},'' {\em Automated Software Engineering},
  vol.~23, pp.~501--532, sep 2016.

\bibitem{Tsantalis2011}
N.~Tsantalis and A.~Chatzigeorgiou, ``{Ranking refactoring suggestions based on
  historical volatility},'' {\em Proceedings of the European Conference on
  Software Maintenance and Reengineering, CSMR}, pp.~25--34, 2011.

\bibitem{Morozoff2010}
E.~P. Morozoff, ``{Using a line-of-code Metric to understand software
  rework},'' {\em IEEE Software}, vol.~27, pp.~72--77, jan 2010.

\bibitem{TieYan2009}
T.-Y. Liu, ``Learning to rank for information retrieval,'' {\em Foundations and
  Trends in Information Retrieval}, vol.~3, no.~3, pp.~225--331, 2009.

\bibitem{Pruijt2017}
L.~Pruijt, C.~K{\"{o}}ppe, J.~M. van~der Werf, and S.~Brinkkemper, ``{The
  accuracy of dependency analysis in static architecture compliance
  checking},'' in {\em Software - Practice and Experience}, vol.~47,
  pp.~273--309, John Wiley and Sons Ltd, feb 2017.

\bibitem{Runeson2009}
P.~Runeson and M.~H{\"{o}}st, ``{Guidelines for conducting and reporting case
  study research in software engineering},'' {\em Empirical Software
  Engineering}, vol.~14, no.~2, pp.~131--164, 2009.

\bibitem{VanSolingen2002}
R.~van Solingen, V.~Basili, G.~Caldiera, and H.~D. Rombach, ``{Goal Question
  Metric (GQM) Approach},'' in {\em Encyclopedia of Software Engineering},
  2002.

\bibitem{David1963}
H.~A. David, {\em The method of paired comparisons}, vol.~12.
\newblock London, 1963.

\bibitem{Perezortiz2017}
M.~Perez-Ortiz and R.~K. Mantiuk, ``A practical guide and software for
  analysing pairwise comparison experiments,'' 2017.

\bibitem{Mikhailiuk2020}
A.~Mikhailiuk, C.~Wilmot, M.~Perez-Ortiz, D.~Yue, and R.~Mantiuk, ``Active
  sampling for pairwise comparisons via approximate message passing and
  information gain maximization,'' in {\em 2020 IEEE International Conference
  on Pattern Recognition (ICPR)}, Jan 2021.

\bibitem{Fix1989}
E.~Fix and J.~L. Hodges, ``Discriminatory analysis. nonparametric
  discrimination: Consistency properties,'' {\em International Statistical
  Review/Revue Internationale de Statistique}, vol.~57, no.~3, pp.~238--247,
  1989.

\bibitem{Herbrich2006}
R.~Herbrich, T.~Minka, and T.~Graepel, ``Trueskill™: A bayesian skill rating
  system,'' in {\em Proceedings of the 19th International Conference on Neural
  Information Processing Systems}, NIPS'06, (Cambridge, MA, USA), p.~569–576,
  MIT Press, 2006.

\bibitem{Wienss2013}
J.~Wienss, M.~Stein, and R.~Ewald, ``{Evaluating Simulation Software Components
  with Player Rating Systems},'' 2013.

\bibitem{Meyerovich2012}
L.~A. Meyerovich and A.~Rabkin, ``{How not to survey developers and
  repositories: Experiences analyzing language adoption},'' in {\em SPLASH
  2012: PLATEAU 2012 - Proceedings of the 2012 ACM 4th Annual Workshop on
  Evaluation and Usability of Programming Languages and Tools}, pp.~7--16,
  2012.

\bibitem{Fleiss1971}
J.~L. Fleiss, ``Measuring nominal scale agreement among many raters.,'' {\em
  Psychological bulletin}, vol.~76, no.~5, p.~378, 1971.

\bibitem{Sas2021}
D.~Sas, P.~Avgeriou, I.~Pigazzini, and F.~{Arcelli Fontana}, ``{On the relation
  between architectural smells and source code changes},'' {\em Journal of
  Software: Evolution and Process}, vol.~34, no.~1, 2021.

\bibitem{Ke2017}
G.~Ke, Q.~Meng, T.~Finley, T.~Wang, W.~Chen, W.~Ma, Q.~Ye, and T.-Y. Liu,
  ``Lightgbm: A highly efficient gradient boosting decision tree,'' in {\em
  Advances in Neural Information Processing Systems} (I.~Guyon, U.~V. Luxburg,
  S.~Bengio, H.~Wallach, R.~Fergus, S.~Vishwanathan, and R.~Garnett, eds.),
  vol.~30, Curran Associates, Inc., 2017.

\bibitem{Stone1974}
M.~Stone, ``Cross-validatory choice and assessment of statistical
  predictions,'' {\em Journal of the royal statistical society: Series B
  (Methodological)}, vol.~36, no.~2, pp.~111--133, 1974.

\bibitem{Wang2018}
X.~Wang, C.~Li, N.~Golbandi, M.~Bendersky, and M.~Najork, ``The lambdaloss
  framework for ranking metric optimization,'' in {\em Proceedings of the 27th
  ACM international conference on information and knowledge management},
  pp.~1313--1322, 2018.

\bibitem{Soliman2018}
M.~Soliman, A.~Rekaby~Salama, M.~Galster, O.~Zimmermann, and M.~Riebisch,
  ``Improving the search for architecture knowledge in online developer
  communities,'' in {\em 2018 IEEE International Conference on Software
  Architecture (ICSA)}, pp.~186--18609, 2018.

\bibitem{Lethbridge2005}
T.~C. Lethbridge, S.~E. Sim, and J.~Singer, ``{Studying Software Engineers:
  Data Collection Techniques for Software Field Studies},'' {\em Empirical
  Software Engineering 2005 10:3}, vol.~10, pp.~311--341, sep 2005.

\bibitem{Tan2021}
J.~Tan, D.~Feitosa, and P.~Avgeriou, ``Do practitioners intentionally self-fix
  technical debt and why?,'' in {\em 2021 IEEE International Conference on
  Software Maintenance and Evolution (ICSME)}, pp.~251--262, 2021.

\bibitem{Maldonado2017}
E.~d.~S. Maldonado, R.~Abdalkareem, E.~Shihab, and A.~Serebrenik, ``An
  empirical study on the removal of self-admitted technical debt,'' in {\em
  2017 IEEE International Conference on Software Maintenance and Evolution
  (ICSME)}, pp.~238--248, IEEE, 2017.

\bibitem{Zampetti2021}
F.~Zampetti, G.~Fucci, A.~Serebrenik, M.~D. Penta, A.~S. aserebrenik, and tuenl
  Massimiliano Di~Penta, ``Self-admitted technical debt practices: a comparison
  between industry and open-source,'' {\em Empirical Software Engineering},
  vol.~26, p.~22, 2021.

\bibitem{Palinkas2015}
L.~A. Palinkas, S.~M. Horwitz, C.~A. Green, J.~P. Wisdom, N.~Duan, and
  K.~Hoagwood, ``{Purposeful sampling for qualitative data collection and
  analysis in mixed method implementation research},'' {\em Administration and
  policy in mental health}, vol.~42, p.~533, sep 2015.

\bibitem{Glaser2017}
B.~G. Glaser and A.~L. Strauss, {\em Discovery of grounded theory: Strategies
  for qualitative research}.
\newblock Routledge, 2017.

\bibitem{Boeije2002}
H.~Boeije, ``A purposeful approach to the constant comparative method in the
  analysis of qualitative interviews,'' {\em Quality \& Quantity}, vol.~36,
  pp.~391--409, 2002.

\bibitem{Glaser1968}
B.~G. Glaser, A.~L. Strauss, and E.~Strutzel, ``The discovery of grounded
  theory; strategies for qualitative research,'' {\em Nursing research},
  vol.~17, no.~4, p.~364, 1968.

\bibitem{Mathison2005}
S.~Mathison, ``Constant comparative method,'' {\em Encyclopedia of evaluation},
  vol.~1, p.~0, 2005.

\bibitem{Sharma2016}
T.~Sharma, ``{ Designite - A Software Design Quality Assessment Tool},'' May
  2016.

\bibitem{Bruch2021}
S.~Bruch, ``An alternative cross entropy loss for learning-to-rank,'' in {\em
  Proceedings of the Web Conference 2021}, pp.~118--126, 2021.

\bibitem{Robillard2010}
M.~Robillard, R.~Walker, and T.~Zimmermann, ``{Recommendation systems for
  software engineering},'' {\em IEEE Software}, vol.~27, pp.~80--86, jul 2010.

\bibitem{Happel2008}
H.-J. Happel and W.~Maalej, ``Potentials and challenges of recommendation
  systems for software development,'' in {\em Proceedings of the 2008
  International Workshop on Recommendation Systems for Software Engineering},
  RSSE '08, (New York, NY, USA), p.~11–15, Association for Computing
  Machinery, 2008.

\bibitem{Palma2012}
F.~Palma, H.~Farzin, Y.-G. Guéhéneuc, and N.~Moha, ``Recommendation system
  for design patterns in software development: An dpr overview,'' in {\em 2012
  Third International Workshop on Recommendation Systems for Software
  Engineering (RSSE)}, pp.~1--5, 2012.

\bibitem{DiRocco2021}
J.~{Di Rocco}, D.~{Di Ruscio}, C.~{Di Sipio}, P.~T. Nguyen, and R.~Rubei,
  ``{Development of recommendation systems for software engineering: the
  CROSSMINER experience},'' {\em Empirical Software Engineering}, vol.~26,
  no.~4, 2021.

\bibitem{Sas2020b}
D.~Sas and P.~Avgeriou, ``{Quality attribute trade-offs in the embedded systems
  industry: an exploratory case study},'' {\em Software Quality Journal},
  vol.~28, pp.~505--534, jun 2020.

\bibitem{Brereton2008}
P.~Brereton, B.~Kitchenham, D.~Budgen, and Z.~Li, ``{Using a protocol template
  for case study planning},'' in {\em Proceedings of the 12th international
  conference on Evaluation and Assessment in Software Engineering}, no.~2006,
  p.~8, 2008.

\bibitem{Lefever2021}
J.~Lefever, Y.~Cai, H.~Cervantes, R.~Kazman, and H.~Fang, ``{On the lack of
  consensus among technical debt detection tools},'' in {\em Proceedings -
  International Conference on Software Engineering}, pp.~121--130, IEEE
  Computer Society, may 2021.

\bibitem{Samarthyam2016}
G.~Samarthyam, G.~Suryanarayana, and T.~Sharma, ``{Refactoring for software
  architecture smells},'' in {\em Proceedings of the 1st International Workshop
  on Software Refactoring - IWoR 2016}, (New York, New York, USA), pp.~1--4,
  ACM Press, 2016.

\bibitem{Martini2018b}
A.~Martini, F.~A. Fontana, A.~Biaggi, and R.~Roveda, ``{Identifying and
  Prioritizing Architectural Debt Through Architectural Smells: A Case Study in
  a Large Software Company},'' pp.~320--335, Springer, Cham, sep 2018.

\bibitem{Terra2013}
R.~Terra, L.~F. Miranda, M.~T. Valente, and R.~S. Bigonha, ``Qualitas. class
  corpus: A compiled version of the qualitas corpus,'' {\em ACM SIGSOFT
  Software Engineering Notes}, vol.~38, no.~5, pp.~1--4, 2013.

\bibitem{Tempero2010}
E.~Tempero, C.~Anslow, J.~Dietrich, T.~Han, J.~Li, M.~Lumpe, H.~Melton, and
  J.~Noble, ``The qualitas corpus: A curated collection of java code for
  empirical studies,'' in {\em 2010 Asia Pacific Software Engineering
  Conference}, pp.~336--345, IEEE, 2010.

\bibitem{Wu2018}
W.~Wu, Y.~Cai, R.~Kazman, R.~Mo, Z.~Liu, R.~Chen, Y.~Ge, W.~Liu, and J.~Zhang,
  ``{Software architecture measurement—Experiences from a multinational
  company},'' in {\em Lecture Notes in Computer Science (including subseries
  Lecture Notes in Artificial Intelligence and Lecture Notes in
  Bioinformatics)}, vol.~11048 LNCS, pp.~303--319, Springer Verlag, 2018.

\bibitem{Martini2018}
A.~Martini, E.~Sikander, and N.~Madlani, ``{A semi-automated framework for the
  identification and estimation of Architectural Technical Debt: A comparative
  case-study on the modularization of a software component},'' {\em Information
  and Software Technology}, vol.~93, pp.~264--279, jan 2018.

\bibitem{Verdecchia2020}
R.~Verdecchia, P.~Lago, I.~Malavolta, and I.~Ozkaya, ``{ATDx: Building an
  architectural technical debt index},'' tech. rep., 2020.

\bibitem{Verdecchia2022}
R.~Verdecchia, I.~Malavolta, P.~Lago, and I.~Ozkaya, ``Empirical evaluation of
  an architectural technical debt index in the context of the apache and onap
  ecosystems,'' {\em PeerJ Computer Science}, vol.~8, p.~e833, 2022.

\bibitem{Mayr2014}
A.~Mayr, R.~Plosch, and C.~Korner, ``{A benchmarking-based model for technical
  debt calculation},'' in {\em Proceedings - International Conference on
  Quality Software}, pp.~305--314, IEEE Computer Society, nov 2014.

\end{thebibliography}
\bibliographystyle{ieeetr}

\end{document}